\documentclass[aps, preprintnumbers, tightenlines, nofootinbib, 11pt]{article}
\usepackage{graphicx} % Required for inserting images
\usepackage{xcolor}
\usepackage{amsmath}
\usepackage{amssymb}
\usepackage{amsfonts,empheq,amsbsy,epsfig,wrapfig,caption,float,comment}
\usepackage{authblk}

\begin{document}

\title{Network Renormalization}
\author[1,2,3]{Andrea Gabrielli}
\author[4,5,*]{Diego Garlaschelli}
\author[5]{Subodh P. Patil}
\author[6,7,8]{M. \'Angeles Serrano}

\affil[1]{\footnotesize Dipartimento di Ingegneria Civile, Informatica e delle Tecnologie Aeronautiche, Universit\`a degli Studi `Roma Tre', Via Vito Volterra 62, 00146 - Rome, Italy}
\affil[2]{`Enrico Fermi' Research Center (CREF), Via Panisperna 89A, 00184 - Rome, Italy}
\affil[3]{Istituto dei Sistemi Complessi (ISC) - CNR, Rome, Italy}
\affil[4]{IMT School for Advanced Studies, Piazza S. Francesco 19, 55100 - Lucca, Italy}
\affil[5]{Lorentz Institute for Theoretical Physics, University of Leiden, The Netherlands}
\affil[6]{Departament de Física de la Matèria Condensada, Universitat de Barcelona, Spain}
\affil[7]{Universitat de Barcelona Institute of Complex Systems (UBICS),  Spain}
\affil[8]{ICREA, Barcelona, Spain}
\affil[*]{Corresponding author: \tt{diego.garlaschelli@imtlucca.it}}

\date{\today}

\maketitle

\begin{abstract}
The renormalization group (RG) is a powerful theoretical framework developed to consistently transform the description of configurations of systems with many degrees of freedom, along with the associated model parameters and coupling constants, across different levels of resolution. It also provides a way to identify critical points of phase transitions and study the system's behaviour around them by distinguishing between relevant and irrelevant details, the latter being unnecessary to describe the emergent macroscopic properties. 
In traditional physical applications, the RG largely builds on the notions of homogeneity, symmetry, geometry and locality to define metric distances, scale transformations and self-similar coarse-graining schemes. 
More recently, various approaches have tried to extend RG concepts to the ubiquitous realm of complex networks where explicit geometric coordinates do not necessarily exist, nodes and subgraphs can have very different properties, and homogeneous lattice-like symmetries are absent. 
The strong heterogeneity of real-world networks significantly complicates the definition of consistent renormalization procedures. In this review, we discuss the main attempts, the most important advances, and the remaining open challenges on the road to network renormalization.

\end{abstract}

\section{Introduction\label{sec:intro}}
Understanding large complex systems emerging in nature and society is particularly challenging because of the presence of multiple interacting scales, which implies no natural `preferred' resolution level at which these systems can be comprehensively represented. 
For instance, were one to model the flow of goods and services between businesses, an important question naturally arises: can the relations that characterize this flow be quantitatively represented in a way that maintains its consistency whenever we aggregate firms into economic sectors or countries?  Similarly, when one examines the structure of metabolic networks, one may wonder whether the properties of its connectedness structure also remain invariant under certain coarse-grainings, no matter the species or specimen this metabolic network derives from. Furthermore, when considering the spread of an epidemic on a network, one is interested in understanding whether the actual long-term aggregate outcomes differ from simple compartment-based analyses, given the underlying modeling and parameter estimation uncertainties. All these questions revolve around the notion of the possible underlying irrelevance of the very fine details of a given network and the process it sustains. 

Physicists are intimately familiar with this notion, which underpins our modern understanding of all material phenomena in the universe. \textit{Renormalization} and \textit{renormalizability} is the idea that at large enough temporal and spatial scales, small-scale details average out. Nothing could be more intuitive, as it is the reason why we do not need to know the precise behaviour of all sub-atomic particles to go about our daily lives. Formalizing this, however, took physicists the better part of several decades, and has led to a spectacular series of insights into the universality of certain physical phenomena. 

Although it may be tempting to transpose the lessons learned in the study of physical systems wholesale to complex networks, a number of underlying assumptions have to be critically reexamined. These include assumptions of locality of interactions along with assumptions of an underlying spatial regularity as realized in most materials, where disorder is systematically treated as a small perturbation. Furthermore, upon considering dynamical processes taking place on a given network, compatibility of any given coarse-graining with this dynamics has to be considered as an additional consistency requirement, whether in terms of any given realization, or at the level of the statistical ensemble it may derive from. It is therefore incumbent upon us to critically reexamine each premise that underpins the paradigm of renormalization -- at its core, the simple idea that one can average over small scales in order to understand aggregate behaviour --  and reformulate these in the specific context of complex networks. 

In this review, we revisit the various approaches that apply renormalization techniques to complex networks. We begin by briefly reviewing the basic notions of renormalization and renormalization group (RG) transformations as traditionally applied to physical systems. The latter typically benefit from an embedding in metric spaces, and therefore lend themselves to techniques premised on notions of geometric regularity and locality. We then discuss how real-world networks -- which often lack a manifest geometric embedding and feature heterogeneity and intertwined connectivity at multiple scales -- require an unavoidable rethinking of various aspects of the renormalization program. We then discuss various approaches that have been proposed to date that implement renormalization on networks. We first review a diverse series of attempts that have proposed various network coarse-graining prescriptions and subsequently discuss approaches that, from different initial standpoints, focus explicitly on generalizing the key notions underlying the RG program to complex networks. Each of these approaches represent viable paths towards implementing the RG on networks with arbitrary degrees of heterogeneity and disorder. We conclude with an outlook of the open problems and research questions that still need to be answered.\\

{\bf Renormalization in statistical physics and field theory. }
In physical systems, locality and causality of interactions conspire in such a manner as to limit the influence of very small-scale fluctuations to a handful of parameters describing the system at macroscopic scales. The first steps towards formalizing this was made in 1966, when Leo P. Kadanoff applied the concept of coarse-graining (originally introduced by Paul and Tatiana Ehrenfest) to statistical mechanical systems in order to understand the behavior in the neighborhood of a critical point of a second-order phase transition~\cite{Kadanoff1966}. 

Kadanoff introduced the two foundational concepts of effective, or coarse-grained, `block-spin’ degrees of freedom for models such as the Ising model, where spins are placed at each site of a lattice and can undergo a \textit{phase transition} between disordered and ordered (or magnetized) phases. Moreover, Kadanoff quantified how the interactions between the block-spin variables should scale as one changes the coarse-graining scale, formalized by the notion of the RG~\cite{kadanoff1976}.
The first numerical calculations on a lattice~\cite{PhysRevLett.31.1411} were followed up by more widely applicable variational methods developed by Kadanoff and others~\cite{PhysRevLett.34.1005, BWSouthern_1978, JAN1978461, DENNIJS1978441}. 

The RG represented a breakthrough in statistical mechanics that first provided a real-space method to formally average over the small-scale details of the system in order to define \textit{effective} degrees of freedom and effective interactions between them at any given scale. This construction served as the springboard for the development of RG methods in the context of continuum field theories, introduced by Kenneth Wilson some years later \cite{wilson1974renormalization} (see also \cite{di1969microscopic}). This not only provided a theoretical foundation for understanding scaling behaviour in critical phenomena, it also formalized what was until then a piecemeal approach to dealing with nominally divergent quantities encountered in field theories. These divergences can be traced back to the mathematical idealization of being able to resolve arbitrarily small distances, for which a well defined, if seemingly arbitrary, prescription for their subtraction can be conceived, as initially proposed by Gell-Mann and Low \cite{gell1954quantum}. This initial prescription lacked any firm conceptual foundation until the works of Kadanoff and Wilson. Since then, RG methods have allowed us to develop a `theory of theories’, through which we can understand how disparate physical systems with vastly different microscopic details can nevertheless exhibit universal behaviour close to their critical points \cite{zinn2021quantum,tauber2012renormalization,PELISSETTO2002549,RevModPhys.86.647}.

Over the past decades, the domain of applicability of RG concepts has been extended to diverse domains that span from cosmology at the largest scales to particle physics at the smallest. In the wider context of statistical mechanics and its domains of applicability, these methods have been extended to include out-of-equilibrium systems on homogeneous lattices, where calculations are facilitated by the presence of local and translation-invariant interactions. Important examples  that aren’t necessarily restricted to physical systems include processes that model epidemic dynamics  \cite{DellaMorte}, biological swarms \cite{Cavagna_2017, Cavagna_2019, Cavagna_2023}, neural networks \cite{Koch-Janusz:2017jhf, Li_2018, PhysRevResearch.2.023369, caso2023}, dynamical and directed percolation \cite{APYoung_1975, PJReynolds_1977}, voter models for opinion dynamics\cite{GALAM200066}, and models for synchronization of non-linear oscillators (such as the celebrated Kuramoto model \cite{Hinrichsen00, Dornic2005}) on one-dimensional chains~\cite{PhysRevE.80.036206}, lattices~\cite{PhysRevE.79.051114} and hierarchical trees~\cite{garlaschelli2019synchronization}. Furthermore, renormalization group methods have also found fruitful application when a system can be abstracted as a tensor network \cite{Levin_2007, Evenbly_2015, Bal_2017, Li_2018}, or in purely in terms of its information theoretic content \cite{PhysRevX.10.011037, PhysRevLett.127.240603, PhysRevLett.126.240601, PhysRevLett.126.200601, PhysRevE.104.064106}.

In contrast to equilibrium settings, the relevant ingredients that determine universality classes in non-equilibrium systems are sometimes not known and, generally, no RG calculation is available at higher orders in perturbative treatments. Nonperturbative RG (NPRG) has emerged as a promising alternative~\cite{PhysRevLett.92.195703, PhysRevLett.92.255703, PhysRevLett.95.100601, Canet_2011} to explore non-equilibrium phase transitions and, for example, has been applied to the diffusive epidemic model (DEP) in lattices~\cite{PhysRevE.96.022137}.\\

{\bf Renormalization of complex networks: what challenges? }
RG methods have taught us that large classes of systems exhibit universal features -- a lesson that has proven useful in building bottom-up models for many natural phenomena. However, very rarely in real-world applications do system interactions exhibit the degree of homogeneity captured by a regular lattice representation. More often, the presence of structural disorder and irregular or spatially inhomogeneous interactions can give rise to network patterns characterized by a large degree of topological and geometrical heterogeneity. 
Indeed, the vast majority of real-world networks are characterized by properties such as a very broad distribution of the number of links (\emph{degree}) per node, a short average path length (\emph{small-world} property) even in very large networks, a non-vanishing average local density of triangles (finite \emph{clustering}) even in sparse networks, a modular (possibly hierarchical) \emph{community structure}, and many more~\cite{posfai2016network,squartini2017maximum,newman2018networks,cimini2019statistical}.
This strong topological irregularity makes the extension of the usual RG approach problematic. Additional challenges are posed by the fact that the systems which network analyses aim to model are finite by their very nature, and for which boundary and finite size effects complicate matters, and necessitate an explicit accounting (see for instance \cite{radicchi2008complex, PhysRevE.79.026104, PhysRevE.104.034304, 10100895, doi:10.1073/pnas.2013825118} for notable progress in understanding finite size effects in scaling behavior).

Indeed, RG, as commonly implemented, is defined by three fundamental steps:
\begin{itemize}
\item[(\emph{i})] {\bf definition of coarse-grained variables} either in real space, i.e. by tiling the latter with identical mesoscopic cells or `blocks' containing multiple microscopic units, or in dual $k$-space, i.e. by decomposing such variables in different wavelength components; 
\item[(\emph{ii})] {\bf averaging out, or marginalizing over the finer details} of the system, represented by local fluctuations --  i.e. the properties of either the small-scale cells or the short wavelength modes; 
\item[(\emph{iii})] {\bf renormalization of the couplings and parameters} of the system whose interaction topology is being coarse-grained\footnote{In theories with a continuum of degrees of freedom, this also necessitates a prescription to deal with the infinities that inevitably arise. This prescription is well understood in the context of particle physics and statistical field theory \cite{Itzykson:1989sx, Burgess:2020tbq}, but will not be needed in the present context given the discrete and finite nature of complex networks.}.
\end{itemize}
Once the above three operations are properly defined, they can be iterated, thereby generating the RG flow which is then examined for trajectories and fixed points.

Now, it is important to realize that step (\emph{i}), which is  necessary for the subsequent steps, is generally straightforward for homogeneous physical systems with local  interactions embedded in translationally invariant metric spaces (e.g. regular lattices or Euclidean geometries), because the  notions of length, distance, and neighborhood make those of `identical mesosocopic cells' or `wavelength' trivial (e.g. a $2D$ lattice can be easily coarse-grained onto a reduced lattice with the same symmetry and dimensionality). An even more trivial case is that of systems with all-to-all homogeneous interactions (as idealized in mean-field models), which can immediately be mapped onto another all-to-all interaction structure.
By contrast, step (\emph{i}) already presents a challenge in highly inhomogeneous and small-world networks, as the examples discussed later in this review will clearly illustrate: how does one properly define block-nodes or slow modes in a complex network?
As a related complication, real-world  networks are typically derived by empirically chosen levels of description, often dictated by data limitations. For instance, information about epidemic spreading or economic shock propagation is often collected at an aggregate resolution level (e.g. contacts between entire sub-populations or flows between economic sectors) that is different from the more fundamental one where the actual dynamics is taking place (e.g. infections spreading among individuals and shocks propagating among single firms). Approaches that model the process directly at the aggregate network level do not capture the actual behavior of the system at the underlying finer scales, as the literature on e.g. financial contagion, epidemiology, and neuroscience has clearly illustrated. 
For instance, approaches that tried to model shock transmission during the 2008 financial crisis and epidemic spread during the Covid-19 outbreak at the aggregate network level failed to capture the empirical phenomenology at the underlying microscopic scales~\cite{bardoscia2021physics,soriano2022modeling}.

Similarly, step (\emph{ii}) usually makes use of the intrinsic symmetries of homogeneous lattices, while in heterogenous networks these symmetries are not manifest. 
Put in a different way, in regular lattices and homogeneous spaces one can replace the notion of connectedness with that of proximity induced by metric coordinates: spatially closer nodes are (more likely to be) connected. Indeed, a typical requirement of physical theories in homogeneous Euclidean spaces or regular lattices, for which RG is typically developed, is {\em locality}. This substantially means that interactions are short range so that the coarse-grained theory contains only local fields and their derivatives up to a maximum finite order. Since this requirement builds on the usual metric notion of node coordinates and homogeneous local neighborhood, it is not easily exported to the case of irregular networked space. In a symmetric binary network a possible ``non-metric" choice is to consider interactions only between pair of nodes corresponding to non-zero entries either of the adjacency matrix (nearest-neighbor interactions) or of a maximum power $n$ of it (up to $n^{th}$-neighbor interactions). Another possibility is to use the discrete metric induced by the minimum path length connecting nodes \cite{song2005self}.
Anyway, the heterogeneous distribution of the number of neighbors at a given path distance from a node and the presence of sparse and disordered long-range interactions, which is at the heart of the widespread small-world property, obscure a clear and unique definition of locality in complex networks and different choices can be considered depending on the meaning of the network nodes and edges and the physical motivation of the embedded model.

Finally, the above complications are intertwined with step (\emph{iii}) in a nontrivial way.  
To see this, first note that, already in disordered lattice systems like spin glasses~\cite{binder1986spin}, the interaction topology might be completely regular (e.g. a grid) but the coupling constants may be very heterogeneous. This heterogeneity can be mathematically represented in terms of random variables such that, thanks to the notion of \emph{self-averaging}, averages over their distribution correspond to spatial averages. In this situation, step (\emph{iii}) entails mapping the probability distribution of the fine-scale coupling constants onto an appropriate induced distribution for the coarse-grained ones~\cite{bar2018renormalization}.
In heterogeneous networks, the same conceptual difficulty extends to the fact that the interaction topology (presence/absence of connections) is itself disordered: indeed, virtually all models of complex networks are random graph models, where the structural disorder is modeled via a probability distribution over graphs with sufficiently realistic properties (using, again, a notion of self-averaging). 
The probabilistic nature of network models implies that, under coarse-graining, step (\emph{iii}) requires that an initial probability distribution over fine-grained graphs is mapped onto a different distribution over coarse-grained graphs. 
Importantly, this mapping defines \emph{per se} a renormalization flow, with potentially its own properties and fixed points.

Motivated by the above limitations, various recent attempts in the literature have tried to introduce the missing comprehensive framework for the renormalization of complex networks.
The goal of this program is that of connecting the multiple possible scales of description, across a wide range of resolution levels, of the structure of heterogeneous networks and the processes they sustain.
In the remainder of this review, we summarize various recently introduced methods employing different yet related strategies to achieve this goal.
We will first give an (unavoidably non-exhaustive) account of the diverse attempts that have been put forward to coarse-grain complex networks, and then expand on specific approaches that have tried to extend one or more of the three steps of the renormalization program outlined above, within the context of heterogeneous networks.

\section{Network coarse-graining approaches\label{sec:cg}}
As we mentioned, the lack of lattice-like symmetries in most real-world complex networks makes the identification of consistent coarse-graining schemes quite challenging.  
We try to systematize the main attempts in this direction by grouping them into broad categories, each based on the main concept being exploited.\\

{\bf Methods based on shortest paths.}
One of the most important attempts to network coarse-graining used the length of shortest paths to define distances between nodes, enabling an iterative renormalization method for network structure based on a box-covering transformation reminiscent of methods in fractal geometry~\cite{mandelbrot1982fractal, samsel2023fractaloriginscommunitystructure}. 
In this approach, nodes within a shortest path distance are coarse-grained to form nodes at a larger scale, producing a flow of renormalized network layers~\cite{song2005self}. 
The method unveiled self-similar and fractal features~\cite{song2005self,PhysRevLett.96.018701,Kim_2007, Fronczak_2024}, suggested new growth mechanisms~\cite{song2006origins}, and led to a classification of networks into universality classes~\cite{radicchi2008complex,rozenfeld2010small}. 
However, due to the small-world property, the discrete metric structure of shortest path lengths in a network represents a poor source of length-based scaling factors and cannot disentangle length scales to reveal network symmetries at a fundamental level. 
As a consequence, self-similarity of the degree distribution under this renormalization approach is not accompanied by the scaling of correlations, particularly clustering. 
In addition, real-space RG transformations have been proposed for network models based on the addition of long-range links to an underlying regular lattice with short-range connectivity, for instance to investigate the behavior of the Watts-Strogatz model near its critical point (in the limit where the density of shortcuts tends to zero~\cite{NEWMAN1999341}) and of a scale-free network model on lattices whose original topological properties remain unchanged after transformation~\cite{PhysRevLett.93.168701}.
These methods however require prior knowledge of the node coordinates, or equivalently of the distinction between short-range links (defining the only path lengths actually involved in the coarse-graining) and long-range ones, which is not known or not even defined for real-world networks.
\\

{\bf Spectral methods. }
Alternatively, renormalization techniques based on spectral properties of networks have been developed by merging nodes in a manner that preserves the behavior of specific process. The spectral coarse-graining method was originally devised for random walks on networks, ensuring the preservation of large-scale behavior by maintaining the largest eigenvalues of the stochastic matrix and the corresponding vectors, effectively decimating the fast modes while leaving the slow modes unchanged~\cite{PhysRevLett.99.038701}. This method has been extended to bipartite networks~\cite{10.1063/1.4773823} and applied to synchronization dynamics of coupled oscillators, where the network Laplacian becomes the relevant matrix~\cite{PhysRevLett.100.174104,CHEN20133036,JIA2019925}. Similar spectral coarse-graining ideas have been employed in the study of synchronization in directed networks~\cite{PhysRevE.83.056123} and controllability~\cite{WANG2017168}. The potential of eigenvectors and eigenvalues of the graph Laplacian to develop a field-theoretic RG approach to order-disorder phenomena~\cite{Aygun_2011} was further explored using deterministic hierarchical networks such as the Cayley tree and the diamond lattice for the Gaussian model~\cite{Aygun_2011,PhysRevE.92.022106}. 
These approaches pave the way for a more comprehensive notion of graph renormalization based on Laplacian spectra~\cite{villegas2023}, to which we devote a separate section later in this review.\\

{\bf Topological methods. } 
In parallel to the aforementioned methods, coarse-graining procedures rooted in specific topological properties of networks have been also proposed. Some of these techniques rely on node centrality measures, such as the degree~\cite{PhysRevE.82.011107} or generalized degree~\cite{LONG2018655}, node similarity measures based on neighbor overlap~\cite{wang2018coarse} (including the traditional notions of e.g. \emph{structurally equivalent} nodes in social networks~\cite{lorrain1971structural} and \emph{trophic(ally equivalent) species} in food webs~\cite{dunne2009food}), and network motifs as recurrent patterns throughout the network~\cite{PhysRevE.71.016127}. Special mention is deserved by decimation schemes like the degree-thresholding renormalization~\cite{serrano2008similarity} and the $k$-core decomposition~\cite{alvarez-hamelin2008core}. The first method removes nodes with a degree below a certain threshold to define a hierarchy of nested subgraphs while the second was specifically designed to unfold a network into its sequence of maximal subgraphs, in which every vertex has at least a certain degree. The two methods have proven effective in revealing hierarchical self-similarity in real systems beyond the scale invariance of the degree distribution; in particular, the clustering spectrum exhibits scaling. 
Moreover, various strategies have been proposed based on modularity in the context of multiscale community detection~\cite{5520347}, aligning with the tradition of multilevel graph partitioning algorithms in computer science. Such algorithms encompass coarsening heuristics based on nodes or edge properties, such as the heavy-edge heuristic for graphs with uniform distributions of connections, which merges nodes linked with large weights~\cite{doi:10.1137/S1064827595287997}. Additionally, community-based coarsening schemes collapse groups of highly interconnected vertices, specifically designed for graphs with power-law degree distribution~\cite{1639360}.
Topological methods as based on the assumption that specific structural patterns, such as (hierarchical) community structure or structurally equivalent nodes, do exist in a network. In absence of such patterns, they would not be able to indicate a coarse-graining scheme. This is quite different from the general idea of renormalization in physics, where e.g. one would like to consistently and iteratively coarse-grain a system as simple as a lattice, even in complete absence of mesoscopic properties.\\

{\bf Symmetry-based methods. } Other approaches look for coarse-graining criteria in networks by studying equivalence classes enriched by some notion of process-based symmetry or invariance. The symmetries that characterize the processes on a network, from biology to computer science, are flexible and local~\cite{leifer2022symmetry}. These flexible invariances provide a general principle for defining the building blocks of a system. Specifically, networks for which there are transformations of nodes that leave the so-called \emph{input trees} invariant give place to the emergence of \emph{equitable partitions}~\cite{boldi2002fibrations}. This property has been used to define concepts such as, among others, \emph{lumpability} of ordinary differential equations~\cite{cardelli2017maximal,cardelli2017erode}, \emph{Boolean backward equivalence}~\cite{argyris2023reducing} and \emph{fibrations coloring}~\cite{morone2020fibration}. The importance of these clustering methods resides in the evidence that symmetry-induced equitable partitions allow the reduction of a chain of dynamical variables without altering the overall dynamics and support the synchronized coherent functions of the system~\cite{gili_synchro}. 
While extremely useful as network reduction techniques that identify process-based equivalence classes of nodes, symmetry-based approaches  typically identify a single partition (usually corresponding to the maximal network reduction). As for purely topological approaches, they are therefore not iterative, and can only be applied if symmetries are present in the first place, thereby differing in spirit from the notion of renormalization.\\

{\bf Coarse-graining in engineering and computer science.} Methods to develop scale-down engineering models of the Internet, along with other communication and transportation networks, to create smaller replicas as test beds with comparable performance, have been a recurrent objective in communication technologies~\cite{4016153,10.1145/1290168.1290173,8370840}. Recently, reducing the size of a graph without significantly altering its fundamental properties has also become an important topic in machine learning, where graph embeddings and deep learning techniques, such as Graph Neural Networks, are icreasingly applied to graph-structured data. However, the typically large size of real-world graph data poses challenges for many machine learning tasks. To address this, some research has introduced coarse-graining schemes to obtain a compressed version of the graph, thereby reducing computational load and complexity to levels suitable for practical application~\cite{chen2018harp,loukas2019graph,jin2020graph,fahrbach2020faster,liang2021mile,huang2021scaling,zhang2023network}. Although these methods are similar in spirit to network renormalization coarse-graining operations, they are typically driven by technical purposes and specific tasks, providing limited insight into the fundamental principles governing the multsicale nature of graph-structured data and complex systems.\\

{\bf Information-theoretic approaches. }
From an information-theoretic perspective, strategies have been developed to explore the interplay between scales. The concept of causal emergence, where a higher scale provides a more informative description of the network's connectivity, minimizing uncertainty in node relationships, has been investigated in~\cite{klein2020emergence}. Additionally, for dynamics on a particular graph, multiple coarse-grained descriptions capturing different features of the original process may exist. Information-theoretic measures can aid in assessing how partitioning the state space of a dynamical process on a network influences the projected dynamics~\cite{10.1093/comnet/cnx055}.\\

{\bf Towards principled renormalization approaches. }
It should be noted that most, if not all, of the approaches discussed so far would collapse to the same procedure if applied to regular lattices, because the underlying concepts become equivalent to each other in that case. By contrast, the heterogeneity of real-world networks can break these correspondences and render different approaches inconsistent with each other. For this reason, there is a need to establish more principled foundations of a theoretical framework for the rigorous renormalization of networks and the processes they support. 
This entails going back to the roots of the  renormalization program that is now relatively well understood for systems defined on regular lattices and homogeneous metric spaces, and especially facing the challenge of generalizing the traditional steps (\emph{i}), (\emph{ii}), (\emph{iii}) that we have discussed at the end of the Introduction.
In the next sections, we will review approaches that have tried to make rigorous efforts in this direction, albeit along different routes.  
To facilitate a coherent treatment of these approaches, here we lay some common ground to orient the reader.  

First, one should generalize step (\emph{i}), i.e. the definition of coarse-grained variables, to arbitrary graphs. 
To lay down some notation, it is convenient to introduce the $N_\ell\times N_\ell$ adjacency matrix $\mathbf{A}^{(\ell)}$ representing the network at a given scale of resolution, i.e. a given hierarchical level (or layer) $\ell$, where $N_\ell$ is the number of nodes observed at that level. The case $\ell=0$ may denote the `finest-grained' observable level (i.e. the native level at which network data are available), while $\ell>0$ denotes the $\ell$-th level of iteration of the coarse-graining process (when considering fine-graining approaches, we will also admit `negative levels' $\ell<0$, as explained below). 
In the case of weighted networks, one is also interested in the weight matrix $\mathbf{W}^{(\ell)}$ representing the intensity of links. 
For simplicity, we will mainly consider undirected networks in our discussion. Step (\emph{i}) is already nontrivial, as we have anticipated, and can indeed be carried out in quite different ways. 
Approaches that try to generalize the `real-space' renormalization procedure focus on defining the coarse-grained variables in terms of `block-nodes' that will identify the rows and columns of the adjacency matrix at the next level. For these approaches, the equivalent of `defining a spatial tiling' is the definition of a certain partition $\Omega_\ell$ mapping the $N_\ell$ nodes $\{i_\ell\}_{i_\ell=1}^{N_\ell}$ observed at level $\ell$ onto a smaller set of $N_{\ell+1}$ coarser nodes (or `blocks') $\{i_{\ell+1}\}_{i_{\ell+1}=1}^{N_{\ell+1}}$ at the next level $\ell+1$. Approaches that try to generalize the $k$-space renormalization procedure focus instead on defining an extended notion of `dual space' for arbitrary graphs, possibly via an explicit operator representation of the adjacency matrix that decomposes the latter into fundamental modes.

Second, one needs to redefine step (\emph{ii}), i.e. integrating out the fine details. Practically, in real space this means defining a mapping from $\mathbf{A}^{(\ell)}$ onto $\mathbf{A}^{(\ell+1)}$ (and from $\mathbf{W}^{(\ell)}$ onto $\mathbf{W}^{(\ell+1)}$ if applicable), while in $k$-space it entails defining a similar surjective mapping between successive instances of the dual operator.
A natural choice is that of assuming a \emph{local} mapping, i.e. one where the entry $a^{(\ell +1)}_{i_{\ell +1} j_{\ell +1}}$ of the coarse-grained matrix is chosen to depend only on the entries $\{a^{(\ell )}_{i_{\ell} j_{\ell}}\}_{i_{\ell} \in i_{\ell +1},j_{\ell} \in j_{\ell +1}}$ of the fine-grained matrix that involve the microscopic nodes contained in $i_{\ell +1}$ and $ j_{\ell +1}$, i.e.
\begin{equation}\label{eq:F}
 a^{(\ell +1)}_{i_{\ell +1} j_{\ell +1}} = \mathcal{F}\left( \{a^{(\ell )}_{i_{\ell} j_{\ell}}\}_{i_{\ell} \in i_{\ell +1},j_{\ell} \in j_{\ell +1}}\right).
\end{equation}
As a useful example, it should be noted that at a purely binary level, a link from block $i_{\ell+1}$ to block $j_{\ell+1}$ is present iff, at the underlying level $\ell$, there is at least one link from any node `inside' $i_{\ell+1}$ (that we will denote as $i_\ell\in i_{\ell+1}$) to any node `inside' $j_{\ell+1}$ (similarly denoted as $j_{\ell}\in j_{\ell+1}$).
This coarse-graining step produces the adjacency matrix $\mathbf{A}^{(\ell+1)}$ with the following entries:  
\begin{equation}\label{Deq:cg rule}
    a^{(\ell +1)}_{i_{\ell +1} j_{\ell +1}} = 1 - \prod_{i_{\ell} \in i_{\ell +1}} \prod_{j_{\ell} \in j_{\ell +1}} (1- a^{(\ell )}_{i_{\ell} j_{\ell}}).
\end{equation}
Similarly, one can introduce a local scheme to obtain the coarse-grained matrix $\mathbf{W}^{(\ell+1)}$ from $\mathbf{W}^{(\ell)}$:
\begin{equation}\label{eq:G}
 w^{(\ell +1)}_{i_{\ell +1} j_{\ell +1}} = \mathcal{G}\left( \{w^{(\ell )}_{i_{\ell} j_{\ell}}\}_{i_{\ell} \in i_{\ell +1},j_{\ell} \in j_{\ell +1}}\right).
\end{equation}
Note that the freedom of choosing $\mathcal{F}$ and $\mathcal{G}$ may lead to scheme dependence, which is a natural circumstance in RG theory. Importantly, $\mathcal{F}$ and $\mathcal{G}$ are in general surjective and not invertible: this corresponds to the (sought-for) loss of small-scale information, which also makes the RG a semi-group in general.
The same property will apply to dual-space approaches, via the lossy integration of fast modes. 

Finally, one needs to generalize step (\emph{iii}), i.e. the renormalization of the coupling constants and model parameters. If the network is a single, deterministic graph (e.g. an empirical network) supporting a specific interaction model or process, then this step will depend on the nature and parameters of the interaction itself. In certain cases, e.g. if the process is in one-to-one correspondence with the procedure used to represent the network in dual space, this step might actually follow directly from step (\emph{ii}), as we will discuss in detail in Sec.~\ref{LRG-sec} in the case of a diffusion process. 
By contrast, if the `network' is not deterministic and is rather represented as a random graph model (with its own parameters) generating several possible outcomes already at the native resolution level $\ell=0$, then step (\emph{iii}) also entails renormalizing the parameters of the random graph model, as we will discuss in Secs.~\ref{Geometric-sec} and~\ref{Ensemble-sec}. 
This might already require a separate fixed-point type of argument, solely for the flow of the random graph model parameters and the induced probability distribution over graphs. 
Importantly, in the latter case the RG flow might be statistically inverted, i.e. one might reverse the flow of the probability distribution from higher levels to lower ones. If applied at the native level $\ell=0$, this reverse flow might actually produce fine-grained ($\ell<0$) versions of the network, with a larger number of nodes and, in general, a sparser topology.

We can now briefly anticipate the  various approaches that will be illustrated in the next sections, and how they relate to each other.

\begin{itemize}
    \item In Sec.~\ref{Geometric-sec} we discuss the \emph{geometric renormalization} framework, which is reminiscent of Kadanoff's real-space approach. It implements step (\emph{i}) by assuming the existence of node coordinates in a latent hyperbolic metric space. Given a real-world network, this approach first looks for the most likely hidden node coordinates and then uses the latter to carry out step (\emph{ii}) by defining block-nodes as aggregates of sufficiently close nodes. Via a suitable implementation of step (\emph{iii}), the coarse-grained network is scale-invariant and statistically consistent with the original hyperbolic setting. The technique can be reversed to enable the self-similar growth of networks. This approach retrieves the traditional statistical physics idea of (random) `spatial lattices', while also encoding distinctive hierarchical features and long-range connections in agreement with the topological properties of real-world networks.
    
    \item In Sec.~\ref{LRG-sec} we discuss the \emph{Laplacian renormalization} framework, which is based on a scheme to coarse-grain networks in dual $k$-space, as in Wilson's approach.  The main idea consists in carrying out step (\emph{i}) by decomposing the Laplacian operator (representing an information diffusion process) of a given graph into its eigencomponents. Then, one implements step (\emph{ii}) by integrating out the fast modes (corresponding to the Laplacian eigenvalues larger than a certain threshold induced by a chosen diffusion timescale) and step (\emph{iii}) by rescaling the time units to obtain a coarse-grained Laplacian. The procedure can also be interpreted in real space in terms of coalescence of the subgraphs formed by the nodes that have been mutually reached by diffusion within the chosen timescale. 
    The approach can characterize how diffusion profiles change under coarse-graining and identify scale-invariant regimes.

    \item In Sec.~\ref{Ensemble-sec} we discuss the \emph{multiscale renormalization} framework, which focuses directly on point (\emph{iii}) while being entirely agnostic with respect to the choice of (\emph{i}), i.e. it works for any possible partition of nodes into supernodes, and carries out point (\emph{ii}) exactly as in Eq.~\eqref{Deq:cg rule}. The approach provides a way to define random graph ensembles that are probabilistically consistent, or stable {\em \`a la} L\'evy, across different aggregation schemes and resolution levels. The goal is that of identifying a functional form of the probability of connection for pairs of nodes that is rigorously invariant upon arbitrary node aggregations. This is achieved by finding the exact fixed point of the RG flow in the space of connection functions. Due to its probabilistic character, the multiscale approach is also invertible, thereby providing a coherent procedure for fine-graining networks observed only at an aggregate level.
\end{itemize}

\section{Geometric network renormalization\label{Geometric-sec}}
We begin with an approach based on the idea of restoring spatial locality in complex networks within a geometric framework~\cite{Boguna2021}. The key concept is that the structural heterogeneity and non-locality of real-world networks is compatible with the hypothesis of curved \emph{hyperbolic} embedding spaces. 
Such framework restores the definition of a geometric RG~\cite{GarciaPerez2018,Zheng2021,zheng2024geometric,van2024renormalization}.\\

{\bf Geometric model.} The geometric soft configuration model~\cite{serrano2008similarity,krioukov2010hyperbolic} combines popularity and similarity dimensions, from which hyperbolic space emerges as a natural embedding for their hierarchical and small-world organization. Nodes are positioned on an underlying, or latent, metric space where closer nodes are more likely to form connections. This links the model to the traditions of random geometric graphs in mathematics~\cite{gilbert1961random,penrose2003random} and latent space models for social networks~\cite{hoff2002latent}, where individuals exist within an unobserved social space and are more likely to connect with others who are closer in distance. Analogously, in the geometric soft configuration model, the latent space does not necessarily correspond to a physical space in which some real networks are embedded, such as geography in trade or airport networks, or $3D$ Euclidean space in the brain. Instead, it represents an abstract space where distances encapsulate the information about the attributes that influence the likelihood that two nodes form a connection. 

In the Newtonian version of the model, known as the $\mathbb{S}^D$ model~\cite{serrano2008similarity}, each node $i_0$ (with $i_0=1,\dots,N_0$) at the `fundamental' hierarchical level $\ell=0$ is assigned a popularity variable, as well as coordinates in a similarity space. Popularity is encoded by a hidden degree $\kappa_{i_0}$ drawn from some arbitrary distribution, $\rho(\kappa)$, typically a power law with exponent $-\gamma$. In dimension $D=1$, the similarity space is a one-dimensional sphere with radius $R_0$, chosen to accommodate $N_0$ nodes with constant density, in which each node $i_0$ is assigned an angular coordinate $\theta_{i_0}$ defining angular distances $\Delta\theta_{i_0 j_0}$ between pairs of nodes $i_0, j_0$.
The probability $p_{i_0 j_0}$ that nodes $i_0$ and $j_0$ are connected takes a gravity-law form~\cite{Boguna2020} and defines a maximum-entropy~\cite{squartini2017maximum} ensemble:

\begin{equation}
	p_{i_0 j_0} = \frac{1}{1+\chi_{i_0 j_0}}, \quad \chi_{i_0 j_0}=\frac{(R_0\Delta\theta_{i_0 j_0})^ {\beta}}{(\hat{\mu}_0\kappa_{i_0}\kappa_{j_0})^{\max{(1,\beta)}}},
	\label{eq:ConProb}
\end{equation}
The parameter $\hat{\mu}_0$ sets the average degree $\langle k\rangle_0$, and the individual expected degrees are proportional to the $\kappa$'s of the corresponding nodes. $\beta$ controls clustering in the network's topology, reflecting the triangle inequality in the latent space, and acts akin to an inverse temperature~\cite{Boguna2020,vanderKolk2022}. The model exhibits an anomalous topological phase transition at $\beta=1$~\cite{vanderKolk2022}. In the geometric phase, corresponding to $\beta>1$, clustering is finite~\cite{serrano2008similarity}, whereas it vanishes in the thermodynamic limit of the non-geometric phase, $0\leq\beta\leq1$. However, in the quasi-geometric regime, $0.5\lesssim\beta< 1$, clustering decays to zero very slowly and scales with the system size, with an exponent that approaches zero when $\beta$ approaches 1~\cite{van2024random}. This implies that finite network structures still retain a significant level of clustering, which is comparable to that of some real-world complex networks. Thus, lower temperatures imply mostly short-range links, while higher temperatures balance the likelihood of different link lengths. 

The $\mathbb{S}^D$-model has a purely geometric formulation, the $\mathbb{H}^{D+1}$ model~\cite{krioukov2010hyperbolic}, where the hidden degree $\kappa$ is mapped to a radial coordinate $r$. This transformation converts the connection probability, Eq.~(\ref{eq:ConProb}), into a Fermi-Dirac function depending solely on the distances between nodes in a hyperbolic plane, which emerges as the latent space supporting the network's geometry. By leveraging statistical inference and machine learning techniques, the most likely coordinates of nodes in the latent hyperbolic space can be estimated by maximizing the congruence between the observed network structure and the $\mathbb{S}^D/\mathbb{H}^{D+1}$ model~\cite{boguna2010sustaining}. The embedding of real networks can be obtained by applying the \textit{Mercator} tool~\cite{GarciaPerez2019,jankowski2023d} which then allows for various downstream tasks, such as visualization, analysis, navigation, and renormalization.

This family of models is highly effective in explaining key features of the connectivity of real networks, such as heterogeneous degree distributions~\cite{serrano2008similarity,krioukov2010hyperbolic,gugelmann2012random}, significant clustering~\cite{serrano2008similarity,krioukov2010hyperbolic,gugelmann2012random,candellero2016clustering,Fountoulakis2021}, small-world phenomena~\cite{serrano2008similarity,abdullah2017typical,friedrich2018diameter,muller2019diameter}, percolation characteristics~\cite{serrano2011percolation,fountoulakis2018law}, spectral aspects~\cite{kiwi2018spectral}, and self-similarity~\cite{serrano2008similarity}. The framework has also been expanded to explain preferential attachment in growing networks~\cite{papadopoulos2012popularity}, weighted networks~\cite{allard2017geometric}, bipartite networks~\cite{serrano2012uncovering,kitsak2017latent}, networks driven by complementarity~\cite{budel2023complementarity}, multilayer networks~\cite{kleineberg2016hidden,Kleineberg2017}, networks with community structures~\cite{zuev2015emergence,garcia-perez:2018aa,muscoloni2018nonuniform}, directed networks~\cite{allard2024geometric}, and networks where nodes have associated features~\cite{featuresnetworkgeometry}. \\

\begin{figure}
    \centering
    \includegraphics[width=1\textwidth]{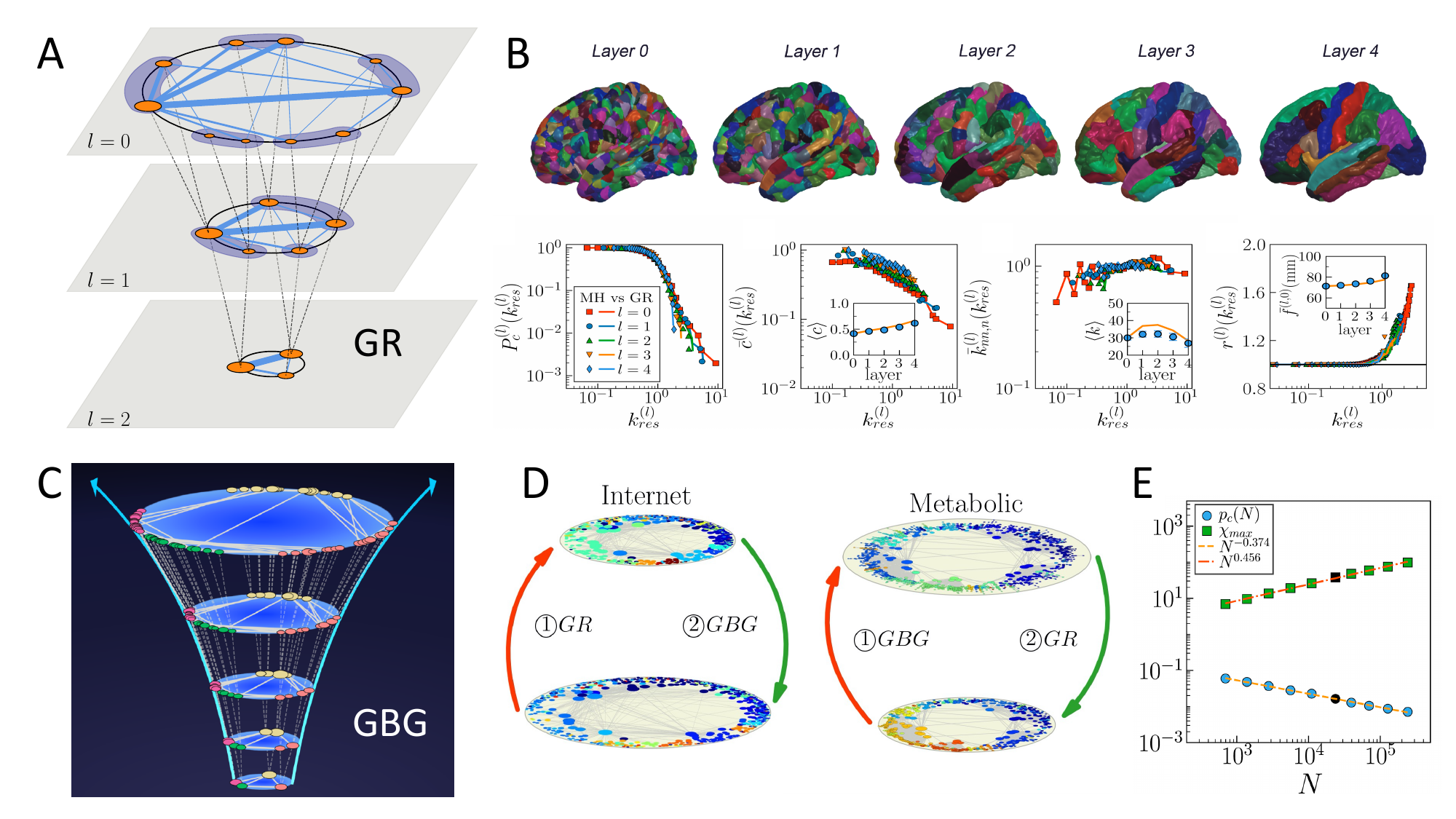}
    \caption{\small \textbf{Geometric network renormalization: direct and inverse methods and results in real networks}.
    {\bf a)} Sketch of the GR transformation. In layer $\ell=0$, non-overlapping blocks of $r=2$ consecutive nodes shaded in gray are defined along the similarity circle. This induces a partition $\Omega_0$ of the original $N_0$ nodes into $N_1=N_0/r$ blocks. The blocks are coarse-grained and represented as supernodes in layer $\ell=1$. Each supernode is assigned an angular coordinate within the region defined by the constituent nodes in $\ell=0$. 
    Finally, two supernodes in $\ell=1$ are connected as described in Eq.~(\ref{Deq:cg rule}). The process can be iterated to produce layer $\ell=2$ from $\ell=1$. Due to the semigroup property, layer $\ell=2$ can also be produced directly from $\ell=0$ by forming blocks of size $r=4$ nodes. {\bf b)} 
    GR replicates the multiscale connectivity structure of human brain connectomes. Multiscale hierarchical representations are reconstructed at five length scales from anatomical data (top): the topological properties (bottom) of the connectomes at each scale are well reproduced by GR applied to the geometric map of the largest resolution layer. {\bf c)} 
    The geometric branching growth process (GBG) evolves networks in a self-similar fashion. {\bf d)} GBG is a statistical inverse of GR, such that the GBG transformation, when applied to a renormalized layer of the Internet, recovers a statistical equivalent to the original Internet network. Similarly, applying GR to a GBG-grown surrogate of the human metabolic network recovers the original architecture. {\bf e)} The combination of GR and GBG offers a way to study critical phenomena using finite size scaling in single-instance real networks, providing a method to estimate the corresponding critical exponents numerically. To illustrate this, results of a bond percolation process mimicking random liks failures in the Internet are shown as a function of the network size, $N_\ell$. 
    The critical bond occupation probability, $p_c$, and the maximum, $\xi_{max}$, of the susceptibility are well fitted by power-laws. The black symbols indicate the original network. 
    The GBG and GR shells were produced with $b=2$. These results suggest a vanishing percolation threshold in the Internet graph, as expected in scale-free networks.}
    \label{fig:GeometricRG}
\end{figure}

{\bf Geometric renormalization technique.} The $\mathbb{S}^1/\mathbb{H}^2$ model lays the grounds for establishing a geometric renormalization (GR) group for complex networks~\cite{GarciaPerez2018,Zheng2021,van2024renormalization,zheng2024geometric}. GR operates iteratively on the hyperbolic map of a real network. 
Once nodes have coordinates in the latent space at level $\ell=0$ (this implements the general step (\emph{i}) illustrated at the end of the Introduction), a new layer $\ell=1$ is produced through coarse-graining and rescaling. 
This involves partitioning the similarity subspace into $N_1$ sectors, via a partition $\Omega_0$ whereby nodes $\{i_0\}\in i_1$ within the same sector $i_1$ (with $i_1=1,\dots,N_1$) become supernodes, which are linked if at least one connection exists between their original nodes as in Eq.~\eqref{Deq:cg rule} (see Fig.~\ref{fig:GeometricRG}),  thereby implementing step (\emph{ii}). Once iterated, this process generates a renormalization flow unfolding networks across multiple scales $\ell\ge 0$ by progressively selecting longer-range connections. GR reveals multiscale self-similarity as a ubiquitous symmetry in real-world networks across different domains~\cite{GarciaPerez2018,van2024renormalization}, and accurately predicts the self-similarity observed in multiscale reconstructions of human brain connectomes, indicating their proximity to the critical structural transition between the small-world and non-small-world regimes~\cite{zheng2019a} (see Fig.~\ref{fig:GeometricRG}). 

These findings are explained by the renormalizability of the $\mathbb{S}^1/\mathbb{H}^2$ model in the geometric $(\beta>1)$~\cite{GarciaPerez2018} and non-geometric $(\beta \le1)$~\cite{van2024renormalization} regimes. The scale transformation for the hidden degrees are obtained by imposing that the connection probability Eq.~(\ref{eq:ConProb}) remains (approximately) invariant. The flow of the angular coordinate can be defined by any transformation that preserves the rotational invariance of the original model. 
Using $z_{i_\ell}=\kappa_{i_\ell}^{\max(1,\beta)}$, the transformations that preserves the probability of connection and, consequently, the topological features of the network in the renormalization flow are
\begin{equation}
	z_{i_{\ell+1}} = \sum_{i_\ell\in i_{\ell+1}}z_{i_\ell},  \quad \theta_{i_{\ell+1}} = \frac{\sum_{i_\ell\in i_{\ell+1}}z_{i_\ell}\theta_{i_\ell}}{\sum_{i_\ell\in i_{\ell+1}}z_{i_\ell}}.
	\label{eq:evolution_kappas}
\end{equation}
with our usual convention of $i_{\ell}\in i_{\ell+1}$ denoting nodes that belong to the supernode $i_{\ell+1}$. The above transformations implement step (\emph{iii}).

The global parameters evolve as $R_{\ell+1}=R_{\ell}/r^\ell$, $\hat\mu_{\ell+1}=\hat\mu_{\ell}/r^{\min(1,\beta)}$, where $r$ is the number of nodes in equally-sized supernodes,  while $\beta$ is scale-invariant. These transformations satisfy the semi-group property: renormalizing twice with groups of $r$ nodes equals renormalizing once with groups of $r^2$. The probability $p_{i_{\ell+1} j_{\ell+1}}$ retains the original form of $p_{i_{\ell} j_{\ell}}$, while the average degree $\langle k\rangle$ is a relevant observable with a flow $\langle k \rangle_{\ell+1} = r^{\nu}\langle k\rangle_{\ell}$ when $\beta> 1$~\cite{GarciaPerez2018}. This also applies to real networks as long as they admit a good embedding. Real networks in the connectivity phase space are shown in Fig.~\ref{fig:GeometricRG}. In phase I, one has $\nu>0$, indicating a flow towards a highly connected graph. In phase II, one has $\nu<0$ and the network flows towards a one-dimensional ring. At the transition between the small-world and non-small-world phases, one has $\nu=0$ and the average degree stays preserved. In phase III, the degree distribution loses its scale-freeness along the flow. Finally, for $\beta\leq 1$, the number of links remains constant under renormalization~\cite{van2024renormalization}.\\

{\bf Extension to weighted networks.} 
The geometric renormalization of weights (GRW)~\cite{zheng2024geometric} produces the multiscale unfolding of a weighted network~\cite{Barrat:2004b,Serrano:2006fu} into a shell of self-similar layers. Application of GR ensures self-similarity of the binarized structure of the network, and the preservation of its weighted structure in the renormalization flow results from imposing the preservation of the relation between the strength and the degree of nodes. The technique is sustained by the renormalizability of the W$\mathbb{S}^D$ model~\cite{allard2017geometric}. In this model, weights between connected nodes in the topology generated by the $\mathbb{S}^D$ model are the result of coupling the weighted structure of the network to the latent metric space. In $D=1$, the transformation of weights, named $\phi$-GRW, is given by a specific choice of the function $\mathcal{G}$ in Eq.~\eqref{eq:G}, namely $w^{(\ell +1)}_{i_{\ell+1} j_{\ell+1}}=C\left[\sum_{i_{\ell} \in i_{\ell +1}} \sum_{j_{\ell} \in j_{\ell +1}} (w^{(\ell )}_{i_{\ell} j_{\ell}})^\phi\right]^{1/\phi}$, which takes the form of a $\phi$-norm where $C$ is a rescaling factor and $\phi$ is a parameter that depends on the weighted and unweighted structure of the network. 
In practice, an effective approximation with practical advantages that retains the semigroup structure property is to select the maximum weight of the renormalized links (see Fig.~\ref{fig:GeometricRG}). This strategy, named sup-GRW, is equivalent to setting $\phi=\infty$ and is effectively reached already for moderate values of $\phi$ due to the fast convergence of the $\phi$-norm of a set of values to the maximum in the set. As an alternative, sum-GRW renormalizes the weights by their sum~\cite{9761989}. This approach is equivalent to setting $\phi=1$ and, in general, does not preserve the relation between hidden strength and hidden degree. Weighted networks with heterogeneous degree distributions from very different domains show geometric scaling when coarse-grained and rescaled using $\phi$-GRW or sum-GRW (see Fig.~\ref{fig:GeometricRG}). \\

{\bf Self-similar fine-graining.} A technique that reverses GR is the Geometric Branching Growth (GBG)~\cite{Zheng2021} model. The GBG model generates self-similar metric expansions of a real network that replicate the original connectivity structure. 
It accurately predicts the self-similar evolution of real-world growing networks, such as the world trade web and the journal citation network, over long time spans. 
To produce a GBG fine-grained layer $\ell-1$ starting from layer $\ell$ and going `backwards' with respect to the coarse-graining direction, every node in the original layer is divided into $r$ descendants with a probability $p$, increasing the population with branching rate $b$. The radius of the similarity subspace is rescaled as $R_{\ell-1}=bR_{\ell}$ to maintain a node density of one. Nodes that do not split retain their coordinates, while each descendant is assigned new values. In principle, GBG allows the generation of arbitrarily large networks.

Two conditions are imposed on the hidden degrees of descendants: first, they must adhere to GR, meaning that their corresponding $z$ values satisfy Eq.(\ref{eq:evolution_kappas}). Second, the distribution of the descendants' hidden degrees $\rho(\kappa)$, or equivalently $\rho(z)$, must preserve that of the ancestor layer. Consequently, $\rho(z)$ must be a stable distribution~\cite{Nolan2018}. By the generalized Central Limit Theorem~\cite{levy1937variables}, stable distributions are the only possible limit distributions for properly normalized and centered sums of {\it i.i.d.} random variables and encompass a rich family of models capable of accommodating fat tails and asymmetry. Concerning angular coordinates, descendants are positioned with slight angular offsets to the left and right of their ancestors, preventing overlaps between descendants of neighboring branching nodes. This transformation retains rotational invariance and the community structure (if present) encoded in the angular distribution of nodes. Once descendant coordinates are assigned, connections between them in the new layer are implemented to ensure the resulting network belongs to the $\mathbb{S}^1$ ensemble. 

Iterative application of GBG, with $\mu_{\ell-1}=b\mu_{\ell}$, produces a sequence of progressively fine-grained self-similar layers, meaning that the original empirical connection probability, degree distribution, clustering coefficient, and community structure are preserved in the flow with the average degree decreasing. Inflationary GBG, with $\mu_{\ell-1}$ readjusted to $a \mu_{\ell-1}$ and $a>1$, avoids it decreasing very fast, or increases it. GBG is a statistical inverse of GR (see Fig.~\ref{fig:GeometricRG}), while inflationary GBG is a statistical inverse of deflationary GR where pruning of links decreases the average degree.

Taken together, the renormalizability of the $\mathbb{S}^1$ model, unweighted and weighted, replicates with high fidelity the multiscale self-similarity observed in real networks. These results indicate that the same principles organize network connectivity at different length scales, simultaneously encoding short- and long-range connections, and GBG results suggest that these principles are also sustained over time. \\

{\bf Applications.} 
 When combined, GR and GBG provide a full up-and-down self-similar multiscale unfolding of a network that covers both large and small scales. Applications include the design of practical and analytic methodologies that exploit the multiscale shell to improve the performance of adapted protocols. For instance, a multiscale navigation protocol that improves single layer results was described in~\cite{GarciaPerez2018}. Scaled-down and scaled-up replicas of real networks~\cite{GarciaPerez2018,Zheng2021}, which preserve their statistical properties including the density of connections, are straightforwardly derived from the geometric renormalization techniques. These replicas allow us to investigate size dependent phenomena in real networks, a prospect that becomes extremely useful in applications such as the study of size-induced stochastic resonance effects, and to explore critical behavior applying finite size scaling techniques~\cite{Zheng2021}. Also, replicas can be used as conveniently reduced testbeds that inform with high fidelity of the qualitative behavior of processes with expensive computational implementation.\\

\section{Laplacian network renormalization}
\label{LRG-sec}
A different general approach to defining a RG for complex networks \cite{villegas2023} uses the concept of Laplacian diffusion among nodes \cite{dedomenico2016spectral,masuda2017random} to induce a run-time detection of the multi-scale intrinsic structures, coarse graining vertices and edges of a generic network, and eventually renormalize a dynamical model defined on it. As for the geometric approach described in the previous section, this renormalization framework can be formulated in an intuitive and physically illustrative real-space version, in strong analogy with the Migdal-Kadanoff RG approach in statistical physics \cite{migdal1976phase, kadanoff1976}. However, its main and defining characteristic is the possibility of formulating RG in the $k$-space version {\em \`a la} Wilson \cite{wilson1974renormalization}. In both cases this framework introduces a recursive procedure for coarsening networks  while preserving their diffusion dynamical features at progressively larger spatio-temporal scales.\\ 

{\bf Laplacian diffusion as neighborhood detector. } 
The first step to formulate the Laplacian RG (LRG) approach for heterogeneous network is to define `equivalent' neighborhoods of different nodes at any fixed arbitrary scale. 
This can be done starting from the operator governing the communication of information in complex undirected networks \cite{newman2010networks,masuda2017random}, i.e. the symmetric Laplacian operator $\mathbf{L}^{(0)}$ that, at the original hierarchical level $\ell=0$, is constructed from the adjacency matrix $\mathbf{A}^{(0)}$ and has entries
\begin{equation}
L^{(0)}_{i_0 j_0}=k_{i_0}\delta_{i_0 j_0}-a^{(0)}_{i_0 j_0},
\label{Laplacian}
\end{equation}
where $\delta_{i_0 j_0}$ is the Kronecker delta symbol and $k_{i_0}=\sum_{j_0=1}^{N_0} a^{(0)}_{i_0 j_0}$ is the degree of node $i_0$ (with $i_0=1,\dots,N_0$). We will consider the case of undirected networks with either binary or weighted connections. 

If $\vec{X}^{(0)}(\tau)$ is any diffusive field defined on the network at time $\tau$ (with $X^{(0)}_{i_0}(\tau)$ the node-$i_0$ component of the field) and governed by the heat diffusion equation $\dot{\vec{X}}^{(0)}(\tau)=-\mathbf{L}^{(0)} \vec{X}^{(0)}(\tau)$, one can formally write
\begin{equation}
    \vec{X}^{(0)}(\tau)=\mathbf{K}^{(0)}(\tau)\vec{X}^{(0)}(0)= e^{-\tau\mathbf{L}^{(0)}}\vec{X}^{(0)}(0),
\end{equation}
where $\mathbf{K}^{(0)}(\tau)\equiv e^{-\tau\mathbf{L}^{(0)}}$ is the diffusion \emph{evolution operator}. 
The element $K^{(0)}_{i_0 j_0}(\tau)$ gives the fraction of field diffused from node $j_0$ to $i_0$ (and viceversa) in a time $\tau$ through all possible paths connecting the two nodes \cite{moretti2019network,Cassi2}. 
Let us call $\{\lambda^{(0)}_{i}\}_{i=1}^{N_0}$ and $\{\vec{\lambda}^{(0)}_{i}\}_{i=1}^{N_0}$ the spectrum of eigenvalues and the corresponding eigenvectors of $\mathbf{L}^{(0)}$, respectively, where for later convenience we assume that the eigenvalues are ordered in increasing order, i.e. $\lambda^{(0)}_1\le\lambda^{(0)}_2\le\dots\le\lambda^{(0)}_{N_0}$. For a connected undirected network, which we consider in this section, all eigenvalues are $\lambda^{(0)}_{i}\ge 0$ with only one null eigenvalue, and the eigenvectors can be chosen to form an orthogonal basis. Clearly, $\mathbf{K}^{(0)}(\tau)$ has the same eigenvectors of $\mathbf{L}^{(0)}$ with eigenvalues $0<e^{-\lambda^{(0)}_{i} \tau}\le 1$.
It is important to note that $\tau$, in analogy with homogeneous spaces (e.g. regular lattices or Euclidean continuous spaces), can be seen as a scale measure on the network: by increasing $\tau$ one can define larger and larger neighborhoods of nodes connected by diffusion. For instance, in a lattice at each $\tau$ there is a corresponding identical spatial scale $l=\sqrt{D\tau}$, where $D$ is the diffusion constant connected to the lattice coordination number, around each node.\\

\begin{figure}[hbtp]
\begin{center}
\includegraphics[width=1.0\columnwidth]{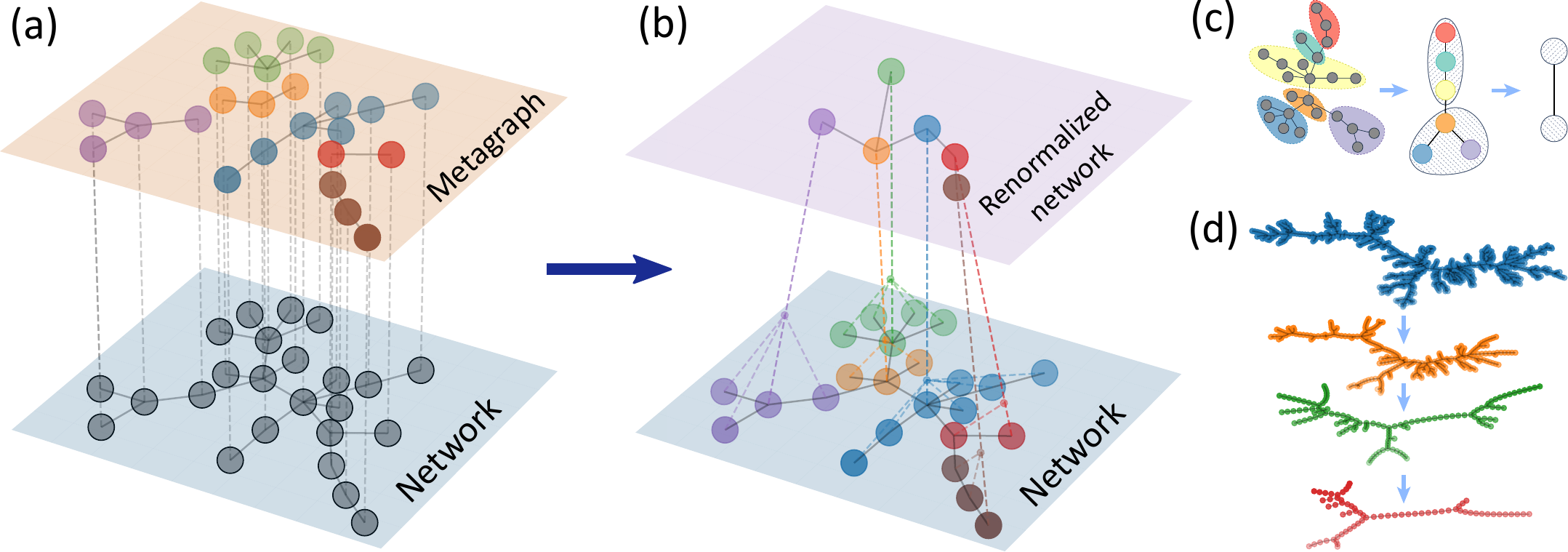}
 \caption{{\bf Real-space construction in the Laplacian network renormalization approach.} {\bf a)} The lower layer $(\ell=0)$ represents the original network $\mathbf{A}^{(0)}$, here a Barab\'asi-Albert network ($N_0=24$, $m=1$), and the upper layer illustrates the partition $\Omega_0$ obtained for $\tau^*=1.96$: different colors identify the $N_1$ Kadanoff supernodes. {\bf b)} Following $\Omega_0$, each block is lumped into a single supernode $i_1$ (with $i_1=1,\dots,N_1$) incident to any edge to the original ones, to create the coarse-grained network $\mathbf{A}^{(1)}$ at the next hierarchical level $\ell=1$. {\bf c)} Illustration of three steps (from left to right, $\ell=0,1,2$) of the real-space LRG process. {\bf d)} LRG coarse-graining of a random tree (from top to bottom, $\ell=0,1,2,3$).}\label{MetaG}
 \end{center}
\end{figure}

{\bf Real-space Laplacian Renormalization Group (LRG). }
Due to its physical meaning, one can use the operator $\mathbf{K}^{(0)}(\tau)$ to partition the $N_0$ nodes of the original ($\ell=0$) network at each arbitrary scale $\tau$ into $N_1$ `diffusionally equivalent' cells (i.e., sub-graphs) which can therefore be coarse-grained into macro-nodes or supernodes to define the network at the next hierarchical level ($\ell=1$). One can then adopt, in analogy with statistical physics, a decimation recipe to set the new {\em renormalized} edges between supernodes starting from the connections at the {\em microscopic} scale, as in Eqs.~\eqref{eq:F} and~\eqref{eq:G}.  In regular lattices, since fixing $\tau$ is equivalent to fixing the spatial scale $l$, this procedure exactly leads to the coarse-graining {\em \`a la} Kadanoff~\cite{Kadanoff1966}.
More specifically, the Laplacian coarse-graining steps can be briefly summarized as follows.
First, define the binarized Laplacian matrix $\boldsymbol{\zeta}^{(0)}(\tau)$ with entries $\zeta^{(0)}_{i_0 j_0}(\tau)=1$ if ${K}^{(0)}_{i_0 j_0}(\tau)\ge {\min[K^{(0)}_{i_0 i_0}(\tau),K^{(0)}_{j_0 j_0}(\tau)]}$ and $\zeta^{(0)}_{i_0 j_0}(\tau)=0$ otherwise. 
Note that ${\zeta}^{(0)}_{i_0 j_0}(0)=\delta_{i_0 j_0}$ (i.e. $\boldsymbol{\zeta}^{(0)}(0)$ is the $N_0\times N_0$ identity matrix $\mathbf{I}^{(0)}$) and $\zeta^{(0)}_{i_0 j_0}(\tau\to\infty)=1$ for all $i_0, j_0$.
At a fixed arbitrary $\tau^*>0$, the matrix $\boldsymbol{\zeta}^{(0)}(\tau^*)$ can be seen as the adjacency matrix of a {\em metagraph} of the original network comprising $N_1$ different connected components which are the clusters of nodes connected by diffusion at the scale $\tau^*$. 
This induces a partition $\Omega_0$ of the original $N_0$ nodes into $N_1$ supernodes.
$\Omega_0$ can then be used to define the adjacency matrix $\mathbf{A}^{(1)}$ of the coarse-grained network, where each supernode $i_1$ (with $i_1=1,\dots,N_1$) corresponds to a diffusion cluster.
Clearly, the larger $\tau^*$ the fewer and larger will such clusters be. For $\tau^*\to\infty$ there will be just a single cluster as diffusion will have connected all network nodes up to have a uniform diffusive field on them. 
To define the coarse-grained weight matrix $\mathbf{W}^{(1)}$ along the lines of Eq.~\eqref{eq:G}, the connections among supernodes can be simply set as weighted edges with a weight equal to the sum of the micro-edges connecting the two cluster at the finer scale, or one can adopt any alternative but consistent decimation procedure in strict analogy with the coarse-graining in statistical physics in order to keep the binary nature of the connections. 
The entire process can then be iterated in order to define higher hierarchical levels $\ell>1$.
This procedure, which corresponds to a generalization, to heterogeneous networks, of the real-space  coarse-graining {\em \`a la} Kadanoff, is illustrated in Fig.~\ref{MetaG}.\\

{\bf $k$-space formulation of the LRG. } 
We now come to the distinctive formulation of the LRG, which instead proceeds in strict analogy with the $k$-space RG approach {\em \`a la} Wilson in statistical field theory \cite{wilson1979problems}. This formulation can indeed be seen as a Fourier-space counterpart of the real-space LRG where the role of the Fourier basis is now played by the set of Laplacian eigenvectors and the role of the wave-vectors by the Laplacian eigenvalues. Indeed in homogeneous spaces, as regular lattices, these eigenvectors coincide with the Fourier basis and the eigenvalues are simply $k^2$. This approach offers a deeper insight into the renormalization process, which has an immediate semigroup structure, and allows one not only to coarse grain the network, but in principle also to appropriately rescale statistical-dynamical models (e.g., Ising model or contact processes) embedded in the network in order to detect and study characteristic (correlation) scales, possible critical points and the system behavior around them. Due to the Hermitian nature of the Laplacian $\mathbf{L}^{(0)}$ and therefore of $\mathbf{K}^{(0)}(\tau)$, it is convenient to adopt the quantum `bra-ket' formalism for the eigenvectors of both. In this way one can rewrite the Laplacian operator in the eigenvectors basis as 
\begin{eqnarray}    
\mathbf{L}^{(0)} = \sum_{{i}=1}^{N_0}\lambda^{(0)}_{i} \left|\lambda^{(0)}_{i}\right>\!\left<\lambda^{(0)}_{i}\right|.
\end{eqnarray}
The completeness of the eigenvector basis ensures that the above representation of the graph Laplacian implements step (\emph{i}) of the general renormalization program illustrated at the end of the Introduction. The subsequent steps are implemented via the following LRG procedure {\em \`a la} Wilson~\cite{villegas2023}:
\begin{itemize}
\item fix a (time) scale $\tau^*$ which will set the new resolution scale (or lower cut-off);
\item partition the Laplacian spectrum into two sets: $\lambda^{(0)}_{i}\ge\lambda^*\equiv 1/\tau^*$, say $n(\tau^*)$ `fast' eigenvalues, and $\lambda^{(0)}_{i}< \lambda^*$, say $N_0-n(\tau^*)$ `slow' eigenvalues;
\item integrate out the `fast' ($\lambda^{(0)}_{i}\ge\lambda^*$) region of the spectrum and redefine the `truncated' Laplacian operator retaining only the contribution of the $N_0-n(\tau^*)$ slow eigenvectors with $\lambda^{(0)}_{i}<\lambda^*$, denoted as $\mathbf{L}^{(0)}_{red}(\tau^*)=\sum_{i=1}^{N_0-n(\tau^*)}\lambda^{(0)}_{i} \left|\lambda^{(0)}_{i}\right>\!\left<\lambda^{(0)}_{i}\right|$. This implements step (\emph{ii}).
\item rescale the diffusion time defining $\tau'=\tau/\tau^*$ so that $\tau^*$ becomes the new time unit. This amounts to a re-definition of the coarse-grained Laplacian to the next hierarchical level $\ell=1$ as $\mathbf{L}^{(1)}\equiv\tau^*\mathbf{L}^{(0)}_{red}(\tau^*)$. This implements step (\emph{iii}).
\end{itemize}
Iterating the above steps produces the renormalized Laplacian $\mathbf{L}^{(\ell)}$ at successive levels ($\ell>1$) of aggregation.

It is important now to underline that the statistical field formulation of the Ising model, contact processes and most of statistical dynamical models characterized by local translational invariant interactions on regular lattices, are characterized by a Gaussian approximation whose kernel is completely determined by the Laplacian operator, which is diagonal in the Laplacian basis (i.e. plane waves in lattices) \cite{Hinrichsen00,Dornic2005}. This remains true if the same models are defined on irregular networks \cite{Burioni1997, Cassi3, PhysRevE.92.022106} with the only difference that the Laplacian operator is no longer translationally invariant. 
Schematically, one can say that the Lagrangian of the theory can be written in terms of the field $\vec{\phi}$ as
\begin{equation}
{\cal L}[\vec{\phi}]=\vec{\phi}\cdot(a\mathbf{I}^{(0)}+\mathbf{L}^{(0)})\vec{\phi}+F[\vec{\phi}]=\sum_{i=1}^{{N}_0}(a+\lambda^{(0)}_{i})|\phi_{\lambda^{(0)}_{i}}|^2+F[\vec{\phi}],
\end{equation}
where $\phi_{\lambda^{(0)}_{i}}=\langle\lambda^{(0)}_{i}|\phi\rangle$, $F[\vec{\phi}]$ is the higher-than-quadratic-order, non-Gaussian part of the Lagrangian, and $\lambda\sim k^2$ with $k$ wave number in the case of regular lattices.
 On the other hand, such Gaussian approximation is the only exactly solvable field theory in all cases, with the non-Gaussian interactions being usually taken into account in the context of perturbation theories. This is one of the key point of Wilson's RG: the Fourier basis, where perturbation theory is developed in field theory, is the one in which the exactly solvable Gaussian approximation is diagonal, i.e. determined by a superposition of independent orthogonal variables. Therefore rescaling is performed in this basis to take advantage of the straightforward reparametrization of the Gaussian approximation, while studying the details of the transformation of the non-Gaussian terms. This is exactly what the Laplacian renormalization approach tries to generalize to the case of non local and heterogeneous networks, seen as embedding spaces for the statistical models.
 
Generalizing what we have already mentioned, the $k$-space LRG scheme can be represented also in real space through the formation of $N_{\ell+1}=N_\ell-n(\tau^*)$ supernodes from the $N_\ell$ original micro-nodes, using the operator $\mathbf{K}^{(\ell)}(\tau)$ through the following steps: (\emph{i}) order the entries $|K^{(\ell)}_{i_\ell j_\ell}(\tau)|$ in descending order; (\emph{ii}) merge micro-nodes into supernodes following this ordered list; (\emph{iii}) stop when the desired number $N_{\ell+1}$ of clusters/supernodes is obtained. Clearly, this is only an approximated real-space representation of the $k$-space coarse-graining procedure. Also in statistical physics the relation between the real space coarse-graining with box cells {\em \`a la} Kadanoff and the $k$-space one, even if in principle possible for any local physical model through special functions, is quite complex and for practical purposes, intractable. Indeed, apart from trivial cases such as the $1D$-Ising model, tractable real-space formulations of the RG, such as the Niemeijer - Van Leeuwen cumulant technique and the Migdal-Kadanoff
 bond-moving decimation approach \cite{kardar2007}, give only approximated results with respect with the Wilson's $k$-space RG formulation coupled to field perturbation theories.\\

{\bf Statistical physics interpretation of the operator $\mathbf{K}^{(0)}(\tau)$. } 
The operator $\mathbf{K}^{(0)}(\tau)$ can be used to build an `intrinsic scales scanner' for networks, possessing all the mathematical properties of the entropic susceptibility or heat capacity for equilibrium statistical physics, so that pronounced maxima (diverging in the infinite size limit) of this quantity detect a sort of phase transitions in the network structural organization.
Indeed the eigenvalues $\{e^{-\lambda^{(0)}_{i} \tau}\}_{i=1}^{N_0}$ of $\mathbf{K}^{(0)}(\tau)$ can be used to define the run-time probability density function 
$\rho(\lambda;\tau)=\sum_{i=1}^{N_0}\delta(\lambda-\lambda^{(0)}_{i})e^{-\lambda\tau}/Z(\tau)$ where $Z(\tau)=\sum_{i=1}^{N_0}e^{-\lambda^{(0)}_{i}\tau}$. This measure has an associated Shannon entropy $S(\tau)=-\sum_{i=1}^{N_0}\rho(\lambda^{(0)}_{i};\tau)\log[\rho(\lambda^{(0)}_{i};\tau)]$.
Since $\mathbf{L}^{(0)}$ for undirected networks is Hermitian and positive semi-definite, $S(\tau)$ can be formally seen also as the Von Neumann entropy $S[\boldsymbol{\rho}(\tau)] = -\operatorname{Tr}[\boldsymbol{\rho}(\tau) \log \boldsymbol{\rho}(\tau)]$, related to the quantum canonical {\em density operator} $\boldsymbol{\rho}(\tau)= {\mathbf{K}^{(0)}(\tau)}/{\mathrm{Tr}[\mathbf{K}^{(0)}(\tau)]}$ \cite{dedomenico2016spectral,ghavasieh2020statistical,InfoCore}, where $\mathbf{L}^{(0)}$ plays the role of the Hamiltonian and $\tau$ that of the inverse temperature \cite{binney1992theory,pathria2011statistical}.  Equivalently, one can say that $Z(\tau)$ is the {\em partition function} related to the free energy $F(\tau)$ by $F(\tau)=-\tau^{-1}\log Z(\tau)$. A direct consequence of this formal analogy is that the quantity 
\begin{equation}
    C(\tau)=-\frac{dS(\tau)}{d\log \tau}
    \label{SHeat}
\end{equation}
has the mathematical properties of an entropic susceptibility, i.e. a heat capacity \cite{InfoCore,villegas2023}. Consequently, a divergence of $C(\tau)$ in the infinite number of nodes limit at a certain scale $\tau^*$, or a pronounced peak in the finite but large $N$ case, can be interpreted as a phase-transition point. This means that $\tau^*$ is an intrinsic network scale at which there is a structural transition, i.e. the topology of the network changes abruptly, similarly to what happens at the correlation length scale in the Ising model. For instance, in a stochastic block model \cite{Holland1983}, characterized by densely connected blocks of the same size with weaker inter-block connections, one will detect two peaks of $C(\tau)$, one at the diffusion scale of the single blocks and another one at the larger scale of the inter-block typical paths. In this sense $C(\tau)$ can be used as a detector of intrinsic scales which can guide the coarse-graining procedure of a network by singling out the characteristic scales at which the network structure shows topological transitions \cite{Multiscale}. \\

{\bf Topologically scale-invariant networks. }
As shown in \cite{villegas2023}, the entropic susceptibility can be rewritten as $C(\tau)=-\tau^2\frac{d T(\tau)}{d\tau}$, with 
\begin{equation}
T(\tau)\equiv{\mbox Tr}[\boldsymbol{\rho} (\tau)\mathbf{L}^{(0)}]=-\frac{d\log Z(\tau)}{d\tau}=\frac{\int d\lambda \,\lambda\,\omega(\lambda)e^{-\lambda\tau}}{\int d\lambda\, \omega(\lambda)e^{-\lambda\tau}},
\end{equation}
where $\omega(\lambda)=\sum_{i=1}^{N_0}\delta(\lambda-\lambda^{(0)}_{i})/N_0$ is the spectral density of $\mathbf{L}^{(0)}$~\cite{Mcgraw2008}. One can define a network as being topologically scale-invariant if $\omega(\lambda)\sim \lambda^{\gamma}$ with $\gamma=d_s/2-1$ with $d_s$ being the Laplacian spectral dimension \cite{Cassi1}. This is equivalent to saying that $C(\tau)=d_s/2$, i.e. the entropic susceptibility is independent of the scale parameter $\tau$. This is for instance the case of regular lattices or random trees, but also of more complex networks such as hierarchical modular networks \cite{moretti2013griffiths}. Due to the above definition of the LRG, all topologically scale-invariant networks keep the same topology under Laplacian scale transformations and coarse-graining \cite{gabrielli2024} (see also \cite{loures2023laplaciancoarsegrainingcomplex}).\\

{\bf Higher-order generalizations.} 
Quite recently, generalizations of the LRG scheme for networks constructed from simplicial complexes and higher-order interactions have been proposed~\cite{nurisso-2024,cheng2024simplex}.
These approaches use a generalized notion of diffusion formally defined via higher-order Laplacian operators~\cite{reitz2020higher,nurisso-2024,cheng2024simplex}. In the resulting picture, information can flow between simplices of any order $k$ via simplices of any other
order $m$. By studying the properties of this
diffusion via cross-order Laplacians, it is possible to probe the existence of characteristic scales in higher-order networks at each order. Specifically, it is possible to extract a cross-order scale signature in simplicial complexes, showing that in most cases, scale-invariance is found only under the lens of specific orders, suggesting the existence of underlying order-specific processes~\cite{nurisso-2024}.
It is also possible to introduce a simplex path integral and a simplex RG to represent trajectories based on a higher-order propagator, leading to a technique to average out short-range high-order interactions in dual $k$-space, while at the same time coarse-graining in real space to reduce the simplex structure encoding  interactions among arbitrary sets of units~\cite{cheng2024simplex}.

\section{Multiscale network renormalization\label{Ensemble-sec}}
The multiscale network renormalization approach~\cite{garuccio2023multiscale,avena2022inhomogeneous,lalli2024geometry} has the objective of introducing a random graph model that is consistent under arbitrary aggregations of nodes. Its motivation originates from two (related) considerations. On one hand, one acknowledges that, for a given network, different node partitions may be relevant for different reasons (for instance, because the different methods discussed in this review might indicate different coarse-grainings of nodes): one should therefore be ready to model the same system consistently under different possible node aggregations. 
On the other hand, one recognizes that, in real-world network data, the meaning of `nodes' is not always homogeneous: sometimes, while the majority of nodes represents the system's units observed at a certain hierarchical level (e.g., individual firms in a production network), other nodes might represent aggregations at coarser hierarchical levels (e.g. entire sectors, or even countries, each aggregating several firms). 
Both considerations imply that, if one is looking for a unique random graph model of the system under consideration, such model should be applicable coherently (i.e., it should produce consistent probability distributions of graphs) after having aggregated (or disaggregated) nodes arbitrarily. 
This means that the multiscale renormalization approach aims at remaining completely agnostic with respect to step (\emph{i}) of the general procedure outlined at the end of the Introduction: one may `tile' the set of nodes arbitrarily, i.e. introduce any desired partition $\Omega_0$, without relying on any notion of metric or diffusional proximity.
Regarding step (\emph{ii}), in principle the approach allows for different choices but, concretely, the choice in Eq.~\eqref{Deq:cg rule} is made~\cite{garuccio2023multiscale} in light of applications where any connection between constituent nodes (e.g. individual firms) is relevant for defining connections between supernodes (e.g. countries or economic sectors). In any case, the distinctive aspect of the multiscale approach is the implementation of step (\emph{iii}) exactly, as we explain below.\\

\begin{figure}
    \includegraphics[width=0.95\textwidth]{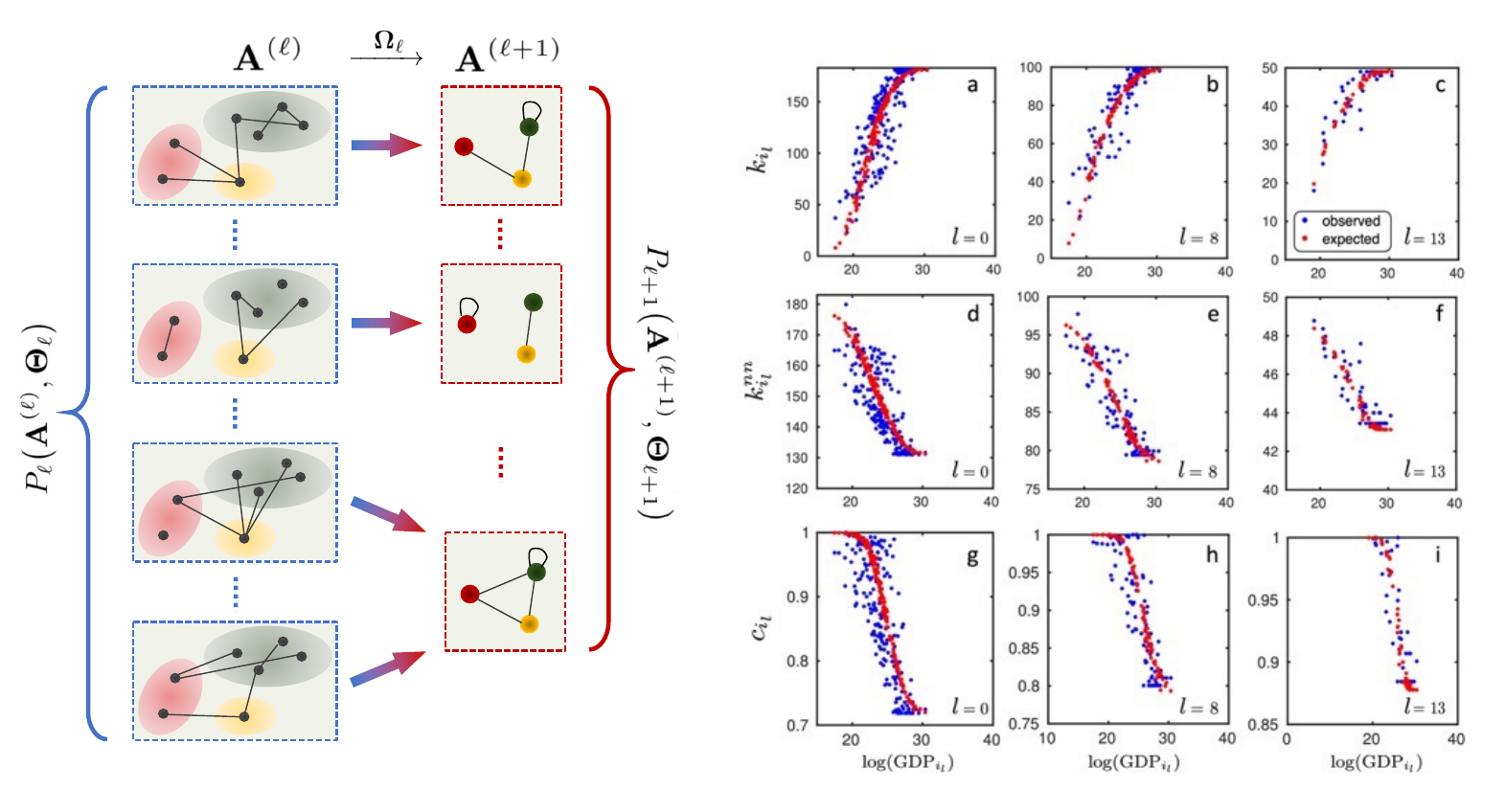}
    \caption{\small \textbf{The multiscale network renormalization approach}. {\bf Left:} Given a probability distribution $P_\ell\big(\mathbf{A}^{(\ell)}|\mathbf{\Theta}_\ell\big)$ of graphs with adjacency matrix  $\mathbf{A}^{(\ell)}$, a node partition ${\Omega}_\ell$ is used to map sets of nodes onto `block-nodes' of the resulting coarse-grained graphs with adjacency matrix $\mathbf{A}^{(\ell+1)}$.
    A link from $i_{\ell+1}$ to $j_{\ell+1}$ is drawn if a link from $i_{\ell}$ to $j_{\ell}$ is present, for any $i_{\ell} \in i_{\ell+1}, j_{\ell} \in j_{\ell+1}$.
    Note that multiple graphs at level $\ell$ may end up in the same graph at level $\ell+1$.
    This coarse-graining will therefore induce a new probability distribution $P_{\ell+1}\big(\mathbf{A}^{(\ell+1)}|\mathbf{\Theta}_{\ell+1}\big)$. The multiscale approach looks for the scale-invariant form of this probability.
    {\bf Right:} Prediction of local topological properties of the renormalized International Trade Network across various levels of geographical aggregation using the multiscale model. (a,b,c): empirical (blue) and expected (red) degree $k_{i_\ell}$ vs $\ln(\mathrm{GDP}_{i_\ell})$ for all $N_\ell$ nodes, for three representative hierarchical levels ($\ell_1=0$, $\ell_2=8$, $\ell_3=13$) such that $N_{\ell_1}=183$ (left), $N_{\ell_2}=100$ (centre) and $N_{\ell_3}=50$ (right). (d,e,f): empirical (blue) and expected (red) average nearest-neighbour degree $k^{nn}_{i_\ell}$ vs $\ln(\mathrm{GDP}_{i_\ell})$ for all $N_\ell$ nodes, for the same three hierarchical levels. (g,h,i): empirical (blue) and expected (red) local clustering coefficient $c_{i_\ell}$ vs $\ln(\mathrm{GDP}_{i_\ell})$ for all $N_\ell$ nodes, for the same three hierarchical levels. Adapted from~\cite{garuccio2023multiscale}.}
    \label{fig:MSM}
\end{figure}

{\bf Invariance under node aggregation. }
The sought-for notion of aggregation-invariant random graphs is similar in spirit to the concept of \emph{stable random variables} in the sense of L\'evy in probability theory~\cite{uchaikin2011chance,samorodnitsky1994m,levy1939addition}, i.e. random variables that remain distributed according to the same law after being combined together (for instance, after taking their sum~\cite{samorodnitsky1994m,levy1939addition} or their maximum~\cite{balkema1977max,gine1990max}). 
Here, using the notation introduced at the end of Sec.~\ref{sec:cg}, one considers a random graph ensemble producing a specific realization $\mathbf{A}^{(\ell)}$ of the graph with probability $P_\ell\big(\mathbf{A}^{(\ell)}|\mathbf{\Theta}_\ell\big)$, where $\mathbf{\Theta}_\ell$ is the set of all parameters of the model. 
Clearly, given the probability distribution $P_\ell\big(\mathbf{A}^{(\ell)}|\mathbf{\Theta}_\ell\big)$ and an arbitrary partition $\Omega_\ell$, a resulting graph ensemble will also be induced at the next level $\ell+1$, described by a certain probability distribution $P_{\ell+1}\big(\mathbf{A}^{(\ell+1)}|\mathbf{\Theta}_{\ell+1}\big)$ (see Fig.~\ref{fig:MSM}).
The key idea of the multiscale renormalization approach~\cite{garuccio2023multiscale} is the identification of a functional form of the graph probability that remains invariant across all hierarchical levels, for all possible partitions (as a result, one can drop the subscript $\ell$ from $P_\ell$). 
Only the (renormalized) parameters $\mathbf{\Theta}_{\ell+1}$ are allowed to depend on the level $\ell+1$, and will in general be related to $\mathbf{\Theta}_\ell$ through ${\Omega}_\ell$. This corresponds to carrying out step (\emph{iii}) via the exact identification of the fixed point of the RG flow in the space of graph probabilities.\\
%%%%%%%%%%%%%%%
%%%%%%%%%%%%%%%

{\bf Multiscale model with independent edges. }
While the most general answer to the above question is currently unknown, it is possible to find the specific solution in the case of graph models with \emph{independent edges}. 
Note that if edges between nodes at a certain level $\ell$ are independent, the same will remain true for edges between blocks of nodes induced at level $\ell+1$ and higher. 
For such models, the full graph probability factorizes in terms of the connection probabilities $p_{i_\ell j_\ell}(\mathbf{\Theta}_\ell)$ between individual pairs of nodes $i_\ell,j_\ell$ as follows:
\begin{equation}\label{eq:factor}
P\big(\mathbf{A}^{(\ell)}|\mathbf{\Theta}_\ell\big)=\prod_{i_\ell,j_\ell}\left[p_{i_\ell j_\ell}(\mathbf{\Theta}_\ell)\right]^{a^{(\ell)}_{i_\ell j_\ell}}\left[1-p_{i_\ell j_\ell}(\mathbf{\Theta}_\ell)\right]^{1-a^{(\ell)}_{i_\ell j_\ell}}.
\end{equation}
The MultiScale Model (MSM)~\cite{garuccio2023multiscale} is defined by the unique nontrivial fixed point of the RG flow in the space of connection probabilities, i.e. the only functional form of the connection probability fulfilling the scale-invariant requirement under the choice in Eq.~\eqref{Deq:cg rule}, i.e.
\begin{equation}\label{eq:cicoria}
p_{i_\ell j_\ell}(\mathbf{\Theta}_\ell)
=\left\{\begin{array}{ll}
1-e^{-\delta x_{i_\ell}x_{j_\ell}f(d_{i_\ell j_\ell})}&i_\ell\ne j_\ell\\
1-e^{-\delta x^2_{i_\ell}f(d_{i_\ell i_\ell})/2}&i_\ell=j_\ell
\end{array}\right.,
\end{equation}
where the parameters $\mathbf{\Theta}_\ell$ have been decomposed into one global parameter $\delta>0$ tuning the overall density of links in the network, a set of $N_\ell$ positive node-specific additive parameters $\{x_{i_\ell}\}_{i_\ell=1}^{N_\ell}$ (called `fitness' values) determining the different individual tendencies of nodes of establishing connections, and (optionally) a set of $N_\ell^2$ dyadic parameters $\{d_{i_\ell j_\ell}\}_{i_\ell,j_\ell=1}^{N_\ell}$ representing node-pair effects such as node-to-node similarity, distance, membership to common communities, etc. 
The function $f(d)$ can be chosen arbitrarily (provided it is monotonic and positive-valued) and can be either increasing or decreasing, depending on whether the dyadic parameters are interpreted as favouring (e.g. similarity) or discouraging (e.g. distance) the establishment of connections, respectively. \\

{\bf Parameter renormalization and aggregation invariance. }
When a node partition $\Omega_\ell$ is used to coarse-grain the network to the next level $\ell+1$, the probability of a coarse-grained configuration $\mathbf{A}^{(\ell+1)}$ will still be described exactly by Eqs.~(\ref{eq:factor}) and~(\ref{eq:cicoria}), with $\ell$ replaced by $\ell+1$ and with  parameters~\cite{garuccio2023multiscale}
\begin{equation}\label{eq:renormparam}
x_{i_{\ell+1}}=\sum_{i_\ell\in i_{\ell+1}}x_{i_{\ell}},\quad d_{i_{\ell+1}j_{\ell+1}}=f^{-1}\left(\frac{\sum_{i_\ell\in i_{\ell+1}}\sum_{j_\ell\in j_{\ell+1}}x_{i_{\ell}}x_{j_{\ell}}f(d_{i_{\ell}j_{\ell}})}{\sum_{i_\ell\in i_{\ell+1}}\sum_{j_\ell\in j_{\ell+1}}x_{i_{\ell}}x_{j_{\ell}}}\right),
\end{equation} 
while $\delta$ is scale-invariant, i.e. it remains unchanged under renormalization (and has therefore no dependence on $\ell$). 
In the special case when $\{d_{i_\ell j_\ell}\}_{i_\ell,j_\ell=1}^{N_\ell}$ are \emph{ultrametric} (i.e. $d_{i_\ell j_\ell}$ can be represented as the distance of nodes $i_\ell$ and $j_\ell$ to their common branching point in a dendrogram where nodes are the leaves), they also become unchanged under renormalization, i.e. $d_{i_{\ell+1}j_{\ell+1}}=f^{-1}\left(f(d_{i_{\ell}j_{\ell}})\right)=d_{i_{\ell}j_{\ell}}$ for any $i_{\ell}\in i_{\ell+1}$ and $j_{\ell}\in j_{\ell+1}$~\cite{garuccio2023multiscale}. 
In another special case, all dyadic effects can be switched off by setting $f(d)\equiv 1$, so that the model becomes entirely fitness-driven. 
Crucially, the fitness parameters themselves cannot be switched off and represent the irreducible features to be considered in the model.

One can rewrite the graph probability in Eq.~(\ref{eq:cicoria}) exactly as 
\begin{equation}
P(\mathbf{A}^{(\ell)},\delta) = \frac{e^{-\mathcal{H}^{(\ell)}_\textrm{eff}\left(\mathbf{A}^{(\ell)},\delta\right)}}{\mathcal{Z}(\delta)}\label{eq:erg}
\end{equation}
where we have introduced the effective Hamiltonian 
\begin{eqnarray}
\mathcal{H}^{(\ell)}_\textrm{eff}(\mathbf{A}^{(\ell)},\delta) &=& -\sum_{i_\ell=1}^{N_\ell} \sum_{j_\ell =1}^{ i_\ell} a^{(\ell)}_{i_\ell,j_\ell} \log{\left[\frac{p_{i_\ell,j_\ell}(\delta)}{1-p_{i_\ell,j_\ell}(\delta)}\right]} \label{eq:Heff}
\end{eqnarray}
and the partition function
\begin{equation}
\mathcal{Z}(\delta) \equiv\sum_{\mathbf{A}^{(\ell)}} e^{-\mathcal{H}^{(\ell)}_\textrm{eff}\left(\mathbf{A}^{(\ell)},\delta\right)}=
e^{\sum_{i_\ell=1}^{N_\ell}\sum_{j_\ell=1}^{N_\ell} x_{i_\ell}x_{j_\ell} f(d_{i_\ell,j_\ell})/2}.
\label{eq:Z}
\end{equation}
Note that the r.h.s. takes the same value irrespective of the level $\ell$ at which it is calculated~\cite{garuccio2023multiscale}, implying that $\mathcal{Z}(\delta)$ is exactly invariant along the renormalization flow as in Kadanoff's real-space renormalization~\cite{kadanoff2000statistical}, confirming the exact fixed-point nature of the connection probability considered.

Unlike the geometric or Laplacian approaches discussed above, the MSM requires neither an embedding metric space with node coordinates, nor a notion of diffusional proximity, in order to guide the renormalization procedure. This can be seen immediately by setting $f(d)\equiv 1$, so that, as already mentioned above, the model only requires an additive fitness attached to nodes. Precisely because there is no notion of vertices being `closer' in a metric or dynamical sense, every partition is allowed by the model, any choice of sets of nodes being aggregated together being admissible. Scale-invariance under this arbitrary notion of aggregation is a more general concept compared with that of proximity-driven scale-invariance, and nontrivially generalizes the ideas of fractality and self-similarity. 

Recently, generalizations of the MSM to weighted~\cite{vvthesis} and directed~\cite{lalli2024geometry} networks have been proposed. The fact that the model does not require geometric distances (which are necessarily symmetric, despite the empirical asymmetry of directed networks) makes the extension to the directed case straightforward, simply via the addition of an additional fitness variable per node. 
With the addition of a third fitness per node, the model can also account for a non-trivial degree of \emph{reciprocity}~\cite{lalli2024geometry}, replicating a widespread property in real-world directed networks~\cite{Garlaschelli2004PatternsOL,garlaschelli2006multispecies,squartini2013reciprocity}. More in general, the fitness variables $x_{i_\ell}$ can be vectors of arbitrary dimension $D$, with the product $x_{i_\ell} x_{j_\ell}$ in the above expressions interpreted as a scalar product~\cite{ialongo2024multiscale,milocco}.\\

{\bf Quenched variant: renormalization of real-world networks. }
In the so-called \emph{quenched} variant of the MSM, the node fitness is interpreted as a deterministic attribute, such as an observable or latent node feature~\cite{garuccio2023multiscale,lalli2024geometry,ialongo2024multiscale,milocco}. 
This variant can therefore model explicitly a real-world network in terms of empirical quantities, as in the family of \emph{fitness models}~\cite{caldarelli2002scale}, or as latent quantities, as in the family of \emph{node embedding} algorithms~\cite{dehghan2022evaluating}. However, while generic models in these families are conceived at a fixed resolution level, the MSM keeps describing the same system consistently via the same connection probabilities at all levels of aggregation, with renormalized parameters. \emph{It is therefore the only additive fitness model that remains a fitness model upon aggregation}~\cite{garuccio2023multiscale,ialongo2024multiscale} and, at the same time, \emph{a method that produces consistent node embeddings across arbitrary coarse-grainings}~\cite{milocco}.

The model turns out to successfully replicate, at several hierarchical levels, the properties of the international trade network~\cite{garuccio2023multiscale,lalli2024geometry} (where the fitness is identified with the empirical Gross Domestic Product of countries, while the dyadic factors are identified with geographic distances described by $f(d)=d^{-1}$) and of inter-firm~\cite{ialongo2024multiscale} and input-output~\cite{milocco} networks (where the out-ward and in-ward fitnesses are defined as the total output and total input of a firm or industry respectively, while $f\equiv1$). Notably, when the dyadic and/or fitness quantities are taken as input from the data, only the global parameter of the model is left as a free parameter. The latter can be fitted at a specific level of resolution, while providing predictions at all other levels. These multiscale predictions are in very good agreement with the empirical network properties at multiple levels~\cite{garuccio2023multiscale,lalli2024geometry,ialongo2024multiscale} (see Fig.~\ref{fig:MSM}).

Note that the model can cope with extremely uneven aggregation schemes where other methods would fail: for instance, in a production network, some nodes may represent individual firms in one EU country, other nodes may represent entire remaining EU countries (with all firms in each country being aggregated into a single node), and yet another `rest of the world' node may represent all the non-EU countries lumped together~\cite{ialongo2024multiscale}.\\

{\bf Annealed variant: clustered scale-free networks and fine-graining.}
In the annealed variant~\cite{garuccio2023multiscale,avena2022inhomogeneous}, the node fitness is considered to be a random variable as in the family of \emph{inhomogeneous random graphs}~\cite{remco}, hence it acts as a `latent' variable with no association to a specific real-world feature.
Here, the requirement of aggregation invariance is applied also to the fitness: at any hierarchical level $\ell$, one should be able to draw the node fitnesses from the same probability density function, without having to `know' their values (or density function) at finer levels. 
This requirement immediately implies that, since the fitness is additive, it should be an $\alpha$-stable random variable~\cite{samorodnitsky1994m}, which means that its density function decays as $x^{-1-\alpha}$ for large $x$. 
To ensure the necessary positivity of the fitness, the exponent should be in the range $\alpha\in(0,1)$, which implies a diverging mean (and all higher moments) for the fitness. The only known $\alpha$-stable distribution in this range is the L\'evy distribution ($\alpha=1/2$), but rigorous results can be derived for any value of $\alpha$ and actually turn out to be largely independent of it~\cite{avena2022inhomogeneous}. 
If nodes are aggregated into blocks of equal size, the fitness retains the same density function, up to a rescaling of global parameters only.
Note that, in its annealed variant, the MSM can generate arbitrarily large networks, and indeed its properties can be studied in the asymptotic limit of a diverging number of nodes~\cite{avena2022inhomogeneous}.

The infinite-mean nature of the fitness makes the expected topological properties of the annealed MSM very different from those produced by similar models with finite-mean fitness. In particular, it can be shown rigorously~\cite{garuccio2023multiscale,avena2022inhomogeneous} that the expected degree distribution $P(k)$ has a universal power-law tail decay as $k^{-2}$, irrespective of the value of $\alpha$. For an aggregation level $\ell$ into $N_\ell$ nodes such that $\delta\sim N_\ell^{-1/\alpha}$, such decay is a pure power-law~\cite{avena2022inhomogeneous}; for coarser aggregations, a density-dependent cutoff emerges in the tail~\cite{garuccio2023multiscale}. Remarkably, the annealed model displays many realistic network properties, including a decaying assortativity profile, a vanishing global clustering coefficient and a non-vanishing local clustering coefficient in the sparse (vanishing density) regime. A non-vanishing local clustering coefficient is not found in other sparse edge-independent models, unless their are assumed to depend on metric distances; therefore the finding that infinite-mean fitness can generate positive local clustering even without geometry is a quite relevant insight of the annealed MSM.

Finally, since $\alpha$-stable random variables are \emph{infinitely divisible} (i.e. they can be expressed as the sum of an arbitrary number of i.i.d. random variables from the same family), in the annealed approach one can fine-grain nodes indefinitely into sub-nodes, each with its own i.i.d. fitness. This means that in this case the renormalization flow defines not only a \emph{semi-group} proceeding bottom-up, but also a \emph{group} proceeding in both directions~\cite{garuccio2023multiscale}.\\

\section{Discussion and future directions}
In this review, we have proposed a critical discussion of the need for a substantial rethinking of renormalization ideas when moving from traditional physical systems endowed with geometry, locality, and homogeneity, to the realm of complex networks where these properties are lacking. We have illustrated the main ideas that have emerged in implementing this rethinking. While some of the resulting approaches have made significant progress towards the introduction of a generalized renormalization framework for heterogeneous networks, several challenges and open questions remain. We list some of these challenges below.\\

{\bf Resolution levels. }
One of the effects of the heterogeneous topology or real-world networks is that understanding the geometric and topological organization into ``functional units" is not as straightforward as in homogeneous lattices or regular trees.
Moreover, it is important to stress that, in several network representations of real-world systems, the available resolution scale defining the nodes of the network is typically not an intrinsic scale of the system, but it is due to observational limitations or to the roughness of data collection. Therefore one would like to determine whether there are characteristic intrinsic scales of the system independently of the observational resolution scale.\\ 

{\bf Renormalizing processes along with network structure. }
One of the future challenges will be that of renormalizing dynamical processes defined on top on networks, along with the underlying graph structure. Some prior attempts in this direction exist in the literature, but have mainly considered processes on regular and/or fractal lattices, for which the structural part of the coarse-graining can be defined quite naturally. 
In particular, exact RG calculations have been applied to Gaussian fields~\cite{10.1143/PTPS.92.108} and random walks~\cite{Burioni_2005} on \emph{fractal networks}, and have been extended to study the fixed points of the Ising model on \emph{Hanoi networks}~\cite{boettcher2011renormalization}, revealing distinct critical non-universal behaviors across different regimes. The critical behavior of percolation on a growing network model has been explored using a decimation RG treatment, showing deviations from the behavior observed in uncorrelated networks~\cite{PhysRevE.67.045102}.
Beyond models, real systems have also been investigated under this lens. For example, a phenomenological coarse-graining procedure has been proposed for activity in networks of neurons. Its application to cells in the hippocampus revealed scaling in both static and dynamic quantities, indicating a nontrivial fixed point in the collective behavior of the network~\cite{PhysRevLett.123.178103}.
In general, we expect that the lack of structural homogeneity in real-world networks and their coarse-grained versions implies a coupling between the renormalization of the dynamical process and that of the topology.
This is one of the future challenges for network renormalization.\\

{\bf Generalized criticality? }
As an important perspective of possible future efforts and progresses in the field, it is important to remark that the strong degree of heterogeneity and non locality of connections in real world networks, in principle not only affect the properties of the transition from local states to critical and collective ones in statistical dynamical models, but it can also give rise to a more complex and chimeric class of behaviors which call even for a more general definition of criticality when the homogeneity of the space of embedding of dynamical models and the locality and homogeneity of interactions are lost. For instance, one can expect that the spreading of an epidemics through a complex network of contacts can manifest more than a simple transition between a collective epidemic state and a localized one. The strong topological heterogeneity and the possible underlying hierarchical structure may determine the appearance of an entire region of the interaction parameters where a critical behavior appear progressively in macroscopic, but partial collective sub-networks. This calls for an extension, induced by topological complexity, of the concept of critical point which is somewhat similar to the one of Griffith phases \cite{Griffiths1969, MAM2010}.\\

{\bf Parameter (ir)relevance: an information-theoretic perspective. }
Coarse graining implies information loss. Far from being a drawback, this loss is the workhorse of RG techniques as it allows us to identify the most \textit{relevant} parameters that describe the behavior of the system at the largest spatial and temporal scales. Renormalization techniques were originally devised to study physical systems with a large number of degrees of freedom and to identify how the parameters describing the system flow from microscopic to macroscopic scales. A crucial aspect of this flow is how some parameters increase in their relevance to the behavior of the system as one goes to larger scales, whereas others become more and more irrelevant. In the context of complex networks, studying this parameter flow from the bottom up can provide us with a first principles understanding of what quantities and control parameters are the most important at the largest spatial and temporal scales. Evidently, a complimentary top-down approach that can also be approached from a purely empirical perspective via the portal of information theory. Specifically, Fisher information and the formalism of information geometry applied to the study of complex networks.

The concept of parameter (ir)relevance in physical systems -- encapsulated by the notion of \textit{power counting renormalizability} in the context of field theories and their hydrodynamic limits -- roughly tracks a derivative expansion in all possible terms that could describe the dynamics of the system. That is, one effectively Taylor expands in theory space, where immediate features become apparent. At large enough scales, terms involving higher derivatives contribute decreasingly to physical observables. From an information theoretic perspective, this implies that the outcome of an ensemble of observations will depend more sensitively on the coefficients of the leading order terms in any derivative expansion. Hence, a direct transcription of notions of parameter relevance are available in terms of information theoretic quantities that moreover, generalizes beyond physical systems. Doing so is not only of immediate conceptual utility, but also opens up a portal towards identifying and understanding the behavior of parameters that determine the behavior of an arbitrary system at the largest scales even in the absence of a microscopic model description or when dealing with arbitrary geometric and ensemble randomness.\\

\begin{figure}[t]
    \includegraphics[width=\textwidth]{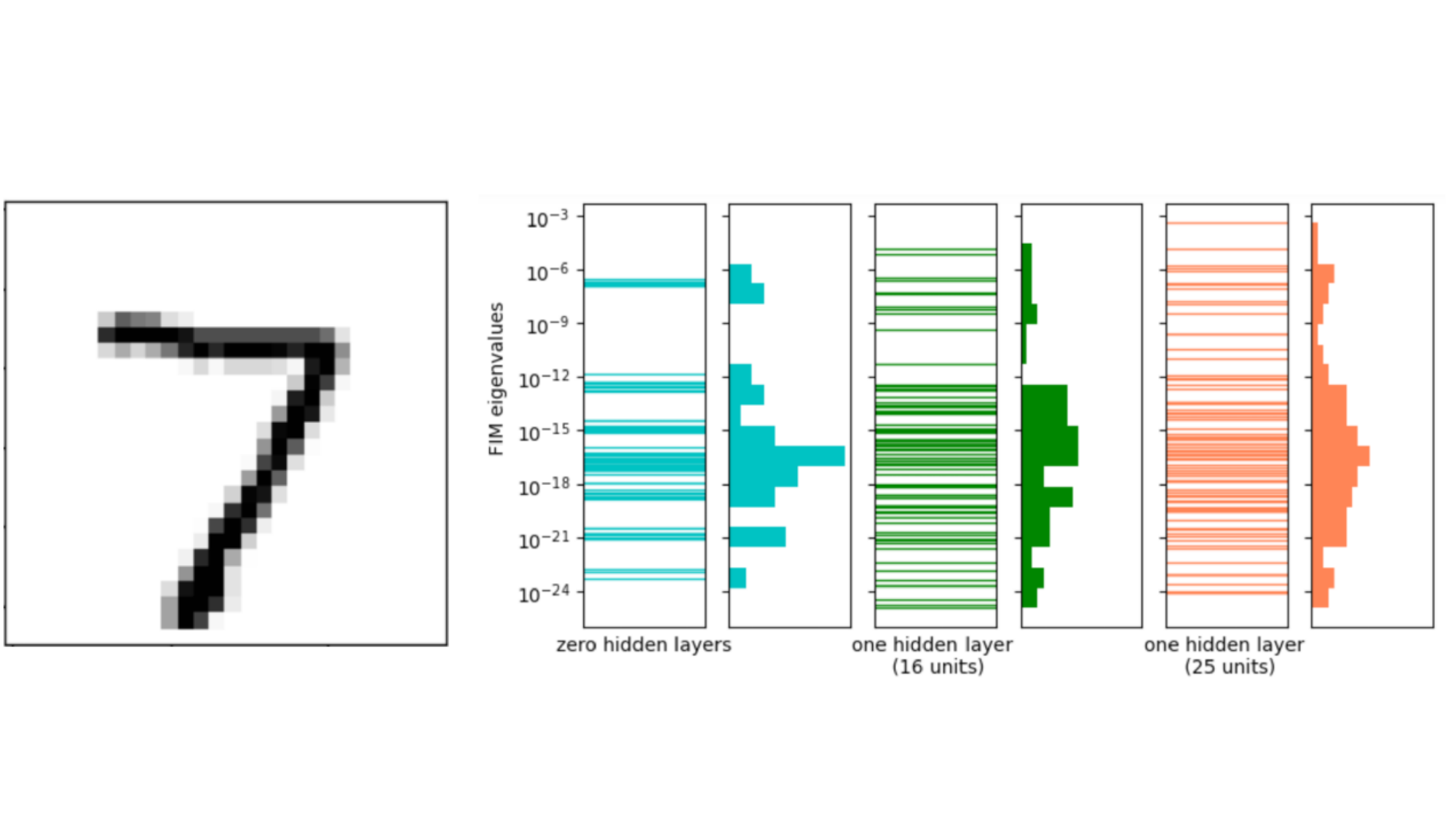}
    \caption{{\bf Information-theoretic parameter flow.} Left: an instance of the handwritten numeral 7. Right: eigenvalues of the Fisher information matrix of 30 trained neural networks that succesfully classified the left figure from the MNIST data set.  An exponential eigenvalue hierarchy is clearly evident, implying that only a handful of collective weights are relevant for the accuracy of the trained network.}
    \label{FIM}
\end{figure}

{\bf Information geometry and parameter flow. }
Moving one step further along the information-theoretic analysis of parameter relevance, given any collection of data one can construct the likelihood function for this data to have been realized by a hypothesized model construction. Maximum-likelihood estimation consists of identifying the parameter set that maximizes this likelihood as being the best inference of what the underlying model might be. How this likelihood varies away from this maximum is
captured by the second derivatives of the log likelihood with respect to the parameters that model it, which very naturally has the structure of a metric that captures distances in parameter space. This geometrization of inference, known as information geometry \cite{amari2016information}, allows one to reimagine statistical (Bayesian) inference and renormalization in a unified framework of parameter flow \cite{Berman:2023rqb}.

When one constructs the Fisher information metric associated with the likelihood constructed from a given data set, one immediately notices a large hierarchy of eigenvalues. This implies that small changes in the  parameters corresponding to the largest eigenvalues will have the largest effect on observed outcomes (see Fig.~\ref{FIM}), the meaning of which maps directly onto the concept of parameter relevance in the RG sense – with the most relevant parameters corresponding to the largest eigenvalues \cite{PhysRevE.98.052112}. It is why given limited access to information about a system, simple models often tend to do unreasonably well in capturing most of its relevant features. In the example depicted in Fig. 4, the collective weights corresponding to the largest eigenvalues of the Fisher information almost entirely dictate the test accuracy of the trained network, with all other parameters and weights being effectively irrelevant. The \textit{Sloppy model} framework \cite{doi:10.1126/science.1238723} consists of identifying these parameters through the hierarchical structure of the eigenvalues of the Fisher information. 

As one successively coarse grains data sets of a given network or simulated realizations of any model embedding, the coarse grained likelihood will result in some eigenvalues increasing in relevance and some decreasing in a manner that maps exactly onto RG relevance. With this realization, one can systematically infer the existence of different phases and control parameters from data alone by studying parameter flow under successive coarse graining of the data, where the only assumptions are the choice of coarse graining scheme, how we parameterize the underlying geometric or ensemble randomness and the priors we might assign to them. This same methodology has been used to identify different phases in heterogeneous chemical systems \cite{2016JSMTE..04.3301H}, and would be a promising avenue to pursue in the context of complex networks.\\

{\bf Outlook. } 
The methods summarized in this review, possibly viewed from the unifying perspective of information theory, promise fundamental advancements towards the foundation of a general theory of scale transformations and RG for complex networks. Each of the above methods faces the central challenges emerging from the intrinsic multiscale nature of complex networks and their behavior under the arbitrary change of resolution scale. Taken together as pieces of a larger puzzle, they delineate some elements of a more general and sought-for framework for the multiscale analysis of irregular networks. At the same time, they already provide fundamental calculation tools to both consistently characterize the intrinsic structural organization of a network at different scales and study the effects of this multiscale architecture on the dynamical processes harboured by the system. From a practical perspective, achieving computational efficiency in network renormalization methods to enable their application to extremely large networks, with hundreds of millions of nodes, remains a significant goal.

The state of the art seems to be mature enough to indicate many new exciting frontiers for research both in network theory and statistical physics, and beyond. We expect future research on network renormalization to have a big impact on different scientific fields where fundamental multiscale models are introduced to capture the architecture and behavior of graph-structured data and dynamical processes taking place on real complex networks. Examples include the dynamics of ecosystems, the relation between structure and function of the human brain, the resilience of socio-economic and financial systems, the spreading dynamics of epidemics, and the compression of graph-structure data for machine learning applications, to name only a few. 

\section*{Authors' contribution}
All authors contributed equally to the design and writing of this review.

\section*{Acknowledgements}
DG acknowledges support from the European Union - NextGenerationEU - National Recovery and Resilience Plan (Piano Nazionale di Ripresa e Resilienza, PNRR), projects `SoBigData.it - Strengthening the Italian RI for Social Mining and Big Data Analytics' (Grant IR0000013, n. 3264, 28/12/2021) and `Reconstruction, Resilience and Recovery of Socio-Economic Networks' (RECON-NET EP\_FAIR\_005 - PE0000013 ``FAIR'' - PNRR M4C2 Investment 1.3). He also acknowledges Luca Avena, Alessio Catanzaro, Elena Garuccio, Rajat Hazra, Leonardo N. Ialongo, Fabian Jansen, Margherita Lalli and Riccardo Milocco for the scientific collaboration on the subject.
AG acknowledges support from the PRIN 2022 PNRR project ``C2T - From Crises to Theory: towards a science of resilience and recovery for economic and financial systems’’ funded by the Italian Ministry of University and Research (MUR) CUP: F53D23010380001. Moreover, he acknowledges Pablo Villegas, Tommaso Gili and Guido Caldarelli for the scientific collaboration on the subject and useful discussions. MAS acknowledges support from TED2021-129791B-I00 funded by MICIU/AEI/10.13039/501100011033 and the ``European Union NextGenerationEU/PRTR''; PID2022-137505NB-C22 funded by MICIU/AEI/10.13039/501100011033 and by ERDF/EU; and \\2021SGR00856 funded by Generalitat de Catalunya. She thanks Mari\'an Bogu\~n\'a, Guillermo Garc\'ia-P\'erez, and Muhua Zheng for the scientific collaboration on the subject. SP wishes to thank Yaneer Bar-Yam, Cliff Burgess, Andrew D. Jackson, Kevin Grosvenor, Leone Luzzatto, and not least, the present coauthors for valuable discussions, collaboration, and exchange of ideas relevant to this review.

\bibliographystyle{naturemag}
\bibliography{bibliography}

\end{document}